\documentclass[a4paper,11pt]{article}
\usepackage{epsfig,lscape}
\usepackage{cite}
\usepackage{color}

\newcommand{\ee}{\mathrm{e}^+ \mathrm{e}^-}
\newcommand{\bb}{\mathrm{b}\overline{\mathrm{b}}}
\newcommand{\cc}{\mathrm{c}\overline{\mathrm{c}}}
\newcommand{\qq}{\mathrm{q}\overline{\mathrm{q}}}
\newcommand{\qb}{\mathrm{b}}
\newcommand{\qc}{\mathrm{c}}
\newcommand{\B}{\mathrm{B}} 
\newcommand{\Bs}{\mathrm{B_s}}
\newcommand{\Bss}{\mathrm{B}^{(*)}_{\mathit{J}}}
\newcommand{\D}{\mathrm{D}} 
\newcommand{\Dplus}{\mathrm{D}^{+}}
\newcommand{\Dzero}{\mathrm{D}^{0}}
\newcommand{\Ds}{\mathrm{D_s}}
\newcommand{\Lambdac}{\mathrm{\Lambda_c}}
\newcommand{\Z}{\mathrm{Z}}
\newcommand{\xE}{\mathit{x}_{\,\mathrm{E}}}
\newcommand{\mxE}{\langle\mathit{x}_{\,\mathrm{E}}\rangle}
\newcommand{\MeV}{\;\mathrm{MeV}}
\newcommand{\GeV}{\;\mathrm{GeV}}
\newcommand{\GeVcc}{\;\mathrm{GeV}/\mathit{c}^{\mathrm{2}}}
\newcommand{\ps}{\;\mathrm{ps} }
\newcommand{\Rb}{\mathrm{R}_\qb}
\newcommand{\Rc}{\mathrm{R}_\qc}

\newcommand{\gbb} {\mathrm {g\to\bb}}
\newcommand{\gcc} {\mathrm {g\to\cc}}

\newcommand {\rphi}{r\phi}
\newcommand {\rz}{rz}
\newcommand{\smcap}[1]{\caption[]{\small \textit{#1}}}
\newcommand{\up}[1]{\raisebox{1ex}[#1][0ex]{\ }}
\newcommand{\dn}[1]{\raisebox{-1.5ex}[0ex][0ex]{#1}}
\newcommand{\etal}{{\rm et al.}}
\newcommand{\EPJ}[3] {Eur.~Phys.~J.\ {C#1} (#2) #3}
\newcommand{\ZPC}[3] {Z.~Phys.\ {C#1} (#2) #3}

\newcommand{\PLB}[3] {Phys.~Lett.\ {B#1} (#2) #3}
\newcommand{\NPB}[3] {Nucl.~Phys.\ {B#1} (#2) #3}
\newcommand{\CPC}[3] {Comp.~Phys.\ {Comm.~#1} (#2) #3}

\newcommand{\NIM}[3] {Nucl.~Instr.\ {Meth.~A#1} (#2) #3}

\newcommand{\PRD}[3] {Phys.~Rev.\ {D#1} (#2) #3}
\newcommand{\JHP}[3] {JHEP\ {#1} (#2) #3}
\newcommand{\JPG}[3] {J.~Phys.\ {G#1} (#2) #3}

\begin{document}
\begin{titlepage}
\begin{center}{\large   EUROPEAN ORGANIZATION FOR NUCLEAR RESEARCH
}\end{center}\bigskip
\begin{flushright}
       CERN-EP/2002-051   \\ 10 July 2002 \\ revised version 26 March 2003
\end{flushright}
\bigskip\bigskip\bigskip\bigskip\bigskip
\begin{center}{\huge\bf  
\boldmath
Inclusive Analysis of the b Quark Fragmentation Function in Z Decays at LEP
\unboldmath
}\end{center}\bigskip\bigskip
\begin{center}{\LARGE The OPAL Collaboration
}\end{center}

\bigskip\bigskip
\bigskip\begin{center}{\large  Abstract}\end{center}

A study of $\qb$ quark hadronisation is presented using inclusively
reconstructed $\B$ hadrons in about four million hadronic $\Z$ decays
recorded in 1992--2000 with the OPAL detector at LEP. The data are
compared to different theoretical models, and fragmentation function
parameters of these models are fitted. The average scaled energy of
weakly decaying B hadrons is determined to be
\[
\mxE = 0.7193 \pm 0.0016 (stat) ^{+0.0038}_{-0.0033}(syst) \ .
\]

\bigskip\bigskip\bigskip\bigskip
\bigskip\bigskip

\begin{center}{\large
(Submitted to Eur.~Phys.~J.~C)
}\end{center}
\end{titlepage}
\begin{center}{\Large        The OPAL Collaboration
}\end{center}\bigskip
\begin{center}{
%begin authorlist PLEASE DO NOT DELETE THIS COMMENT
G.\thinspace Abbiendi$^{  2}$,
C.\thinspace Ainsley$^{  5}$,
P.F.\thinspace {\AA}kesson$^{  3}$,
G.\thinspace Alexander$^{ 22}$,
J.\thinspace Allison$^{ 16}$,
P.\thinspace Amaral$^{  9}$, 
G.\thinspace Anagnostou$^{  1}$,
K.J.\thinspace Anderson$^{  9}$,
S.\thinspace Arcelli$^{  2}$,
S.\thinspace Asai$^{ 23}$
% ...
%end authorlist PLEASE DO NOT DELETE THIS COMMENT
}\end{center}
\bigskip\bigskip
%begin institutes
$^{  1}$School of Physics and Astronomy, University of Birmingham,
Birmingham B15 2TT, UK
\newline
$^{  2}$Dipartimento di Fisica dell' Universit\`a di Bologna and INFN,
I-40126 Bologna, Italy
\newline
$^{  3}$Physikalisches Institut, Universit\"at Bonn,
D-53115 Bonn, Germany
\newline
% ...
$^{  5}$Cavendish Laboratory, Cambridge CB3 0HE, UK
\newline
% ...
$^{  9}$Enrico Fermi Institute and Department of Physics,
University of Chicago, Chicago IL 60637, USA
\newline
% ...
$^{ 16}$Department of Physics, Schuster Laboratory, The University,
Manchester M13 9PL, UK
\newline
% ...
$^{ 22}$Department of Physics and Astronomy, Tel Aviv University,
Tel Aviv 69978, Israel
\newline
$^{ 23}$International Centre for Elementary Particle Physics and
Department of Physics, University of Tokyo, Tokyo 113-0033, and
Kobe University, Kobe 657-8501, Japan
\newline
% ...
%end notes

\section{Introduction}

Hadronisation, the transition of quarks into hadrons, is a strong
interaction phenomenon which cannot yet be calculated from first
principles within QCD. Monte Carlo event generators are used instead
which rely on phenomenological models of this process.  To some extent
these models can be distinguished from each other by the shape of the
predicted hadron energy distribution.  Hadronisation of heavy quarks
leads to a significantly harder hadron energy spectrum than for
lighter quarks~\cite{bib-bjorken}.  Experimentally, heavy quark
hadronisation is of special interest, because in this case the hadron
containing the primary quark can easily be identified.

A precise measurement of the $\B$ hadron\footnote{All hadrons
  containing a $\qb$ quark will be referred to as $\B$ hadrons
  throughout this paper.} energy distribution allows the various
hadronisation models available to be tested, and also helps to reduce
one of the most important systematic uncertainties in many heavy
flavour analyses. Earlier measurements of the $\B$ hadron energy
distribution usually fell into one of three categories. 1) Some
analyses were based on a measurement of the energy distribution of
certain exclusive $\B$ hadron decays, mostly $\B\to\D^{*}\ell\nu$, to
constrain the $\B$ hadron energy as precisely as
possible~\cite{bib-bfralop95,bib-bfraleph01}. However, this leads to
small candidate samples and thus to a large statistical uncertainty.
2) Other analyses attempted to increase the sample size by extracting
the energy distribution of leptons from inclusive $\B\to\ell$
decays.  Unfortunately the modelling of the lepton energy spectrum
introduces large additional systematic
uncertainties~\cite{bib-bfroldlep,bib-bfropal00}. 3) The most precise
results so far have been achieved through the fully inclusive
reconstruction of the $\B$ hadron energy~\cite{bib-bfrsld02}. The
analysis presented here identifies $\B$ hadrons inclusively using
secondary vertices.

\section{Data sample and event selection}
\label{sec-bfrsample}

This analysis uses data taken at or near the $\Z$ resonance with the
OPAL detector at LEP between 1992--2000. A detailed description of the
OPAL detector can be found elsewhere \cite{bib-opaldet}. The most
important components of the detector for this analysis are the silicon
microvertex detector, the tracking chambers, and the electromagnetic
calorimeter. The microvertex detector consisted of two layers of
silicon strip detectors which provided high spatial resolution near
the interaction region. The central jet chamber was optimised for good
spatial resolution in the plane perpendicular to the beam
axis\footnote{The OPAL coordinate system is defined as a right-handed
  Cartesian coordinate system, with the $x$-axis pointing in the plane
  of the LEP collider towards the centre of the ring and the $z$-axis
  along the electron beam direction.}.  The resolution along the beam
direction was improved by the $z$~information delivered by the silicon
microvertex detector (except in the first version present in 1992),
by a vertex drift chamber between the silicon detector and the jet
chamber, and by dedicated $z$-chambers surrounding the other tracking
chambers.  The central detector provided good double track resolution
and precise determination of the momenta of charged particles by
measuring the curvature of their trajectories in a magnetic field of
$0.435\;$T. The solenoid was mounted outside the tracking chambers but
inside the electromagnetic calorimeter, which consisted of
approximately $12\thinspace 000$~lead glass blocks. The
electromagnetic calorimeter was surrounded by a hadronic calorimeter
and muon detectors.

Hadronic events are selected as described in Ref.~\cite{bib-oldrb},
giving a hadronic $\Z$ selection efficiency of $(98.1\pm 0.5)\,\%$ and
a background of less than $0.1\,\%$. Only data that were taken with
the silicon microvertex detector in operation are used for this
analysis.  A data sample of about 3.91 million hadronic events is
selected. This includes 0.41 million events taken for detector
calibration purposes during the years 1996--2000, when LEP was
operating at higher energies.

A total of 23.81 million Monte Carlo simulated events are used: 16.81
million events were generated with the JETSET 7.4
generator~\cite{bib-jetset}, 2 million events were generated with
HERWIG 5.9~\cite{bib-herwig5}, and 5 million events were produced by
HERWIG 6.2~\cite{bib-herwig6}. The JETSET event sample includes 4.93
million $\bb$ events and 3.19 million $\cc$ events in dedicated heavy
flavour samples. All other samples are mixed five flavour $\Z\to\qq$
event samples. The choice of important parameters of the event
generators is described in~\cite{bib-mcsetup}. All Monte Carlo
simulated events are passed through a detailed detector
simulation~\cite{bib-gopal}. The same reconstruction algorithms as for
data are applied to simulated events.

The analysis is performed separately for the data of different years,
where detector upgrades, in particular of the silicon microvertex
detector~\cite{bib-siupgrade}, and recalibrations lead to different
conditions. Separate samples of JETSET Monte Carlo are available for
all years. HERWIG Monte Carlo is only available for the largest
homogeneous dataset taken in 1994, and therefore HERWIG-related
studies are performed exclusively for this dataset.

In the 1993 and 1995 runs, part of the data was taken at
centre-of-mass energies about $1.8\GeV$ above and below the peak of
the $\Z$ resonance. The $\B$ hadron energy distribution is sensitive
to energy losses due to initial state radiation prior to the
annihilation process. Initial state radiation is heavily suppressed at
and just below the $\Z$ resonance, but it has significant impact in
the dataset taken at an energy of $m_{\Z}+1.8\GeV$. The latter samples
are therefore treated separately, with Monte Carlo samples simulated
for the appropriate energy, giving a total of eleven separate data and
JETSET Monte Carlo samples.

\boldmath
\section{Preselection of $\Z\to\bb$ events}
\unboldmath

The thrust axis is calculated for each event using tracks and
electromagnetic clusters not associated with any tracks. To select
events within the fiducial acceptance of the silicon microvertex
detector and the barrel electromagnetic calorimeter, the thrust axis
direction is required to satisfy $|\cos\theta_T|<0.8$, where
$\theta_T$ is the thrust angle with respect to the beam direction.

To achieve optimal $\qb$-tagging performance, each event is forced
into a 2-jet topology using the Durham jet-finding
scheme~\cite{bib-durham}.  In calculating the visible energies and
momenta of the event and of individual jets, corrections are applied
to prevent double counting of energy in the case of tracks with
associated clusters~\cite{bib-chargino172}. A $\qb$-tagging algorithm
is applied to each jet using three independent methods: lifetime tag,
high $p_t$ lepton tag and jet-shape tag. This algorithm was developed
for and used in the OPAL Higgs boson searches. A detailed description
of the algorithm can be found in~\cite{bib-higgs183}. Its
applicability to events recorded at the $\Z$ resonance peak was
already shown in~\cite{bib-yukawa}.  The $\qb$-tagging discriminants
calculated for each of the jets in the event are combined to yield an
event $\qb$ likelihood ${\cal B}_{\rm event}$. Both the jet b-tagging
discriminant and ${\cal B}_{\rm event}$ have values between zero and
one and correspond approximately to the probability of a true $\qb$
jet or $\bb$ event, respectively. For each event, ${\cal B}_{\rm
  event}>0.2$ is required.  The $\Z\to\bb$ event purity is 83\% after
this requirement, and the efficiency is 54\% at this stage.

The $\qb$ hemisphere tag efficiency obtained from Monte Carlo
simulation is compared to the actual value in data using a double tag
approach as described in~\cite{bib-rb}. The efficiencies obtained
this way in both simulation and real data are found to agree to within
5\% in all subsamples. Nevertheless a correction is applied to the
Monte Carlo efficiency to further improve the agreement.

\boldmath
\section{Reconstruction of $\B$ hadron energy}
\unboldmath

\begin{figure}
  \begin{center}
    \epsfig{file=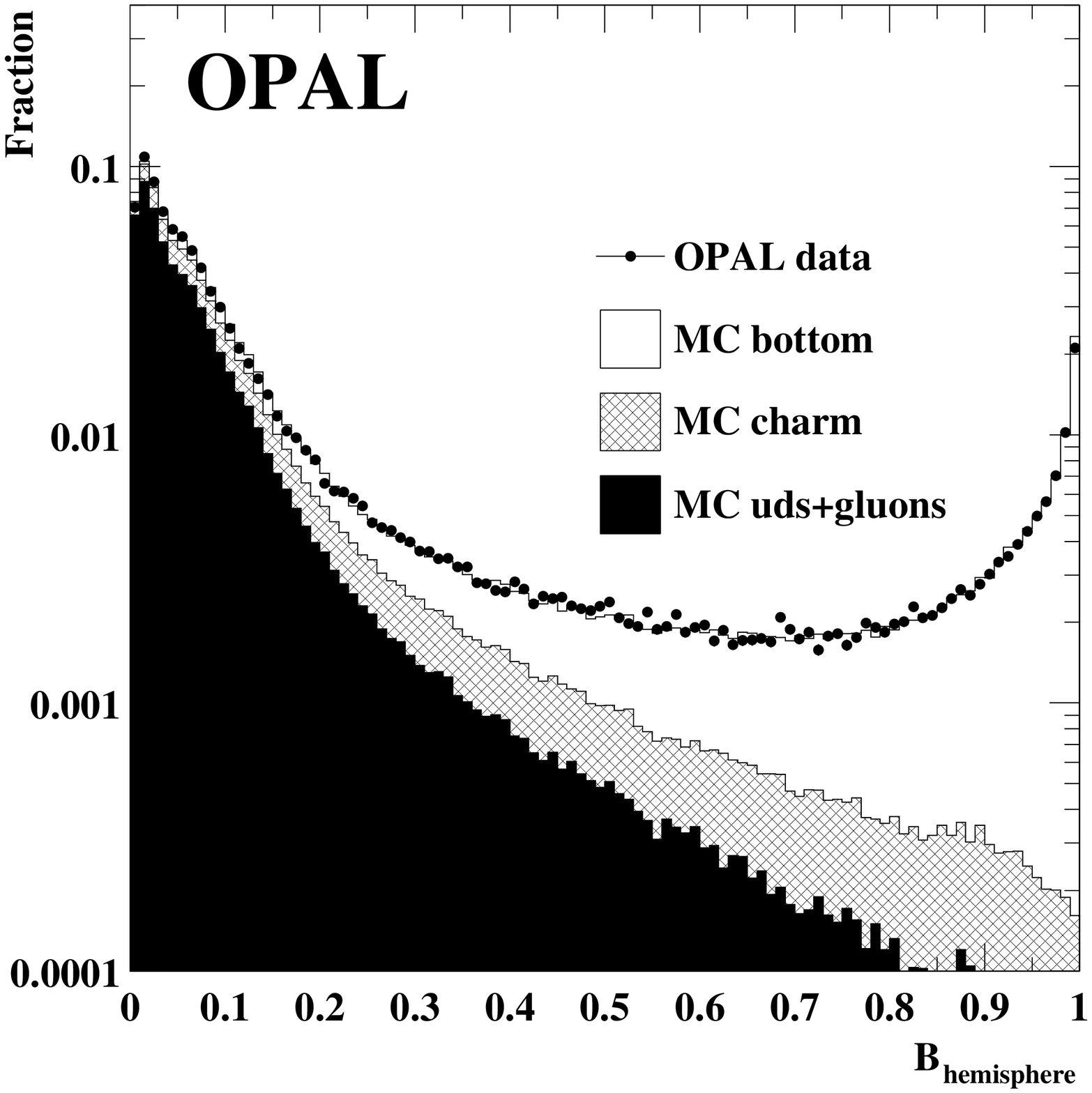, width=10cm}
  \end{center}
  \smcap{\label{fig-lbhemi} Distribution of the jet b-tagging
    discriminant in data (points with error bars) and Monte Carlo
    (histograms). The contributions from $\qb$ jets, $\qc$ jets, and
    light quark or gluon jets are shown as open, hatched, and black
    area respectively.  Jets with a b-tagging discriminant above 0.8
    in a jet in the opposite hemisphere are retained for analysis.}
\end{figure}

The primary event vertex is reconstructed using the tracks in the
event constrained to the average position of the $\ee$ collision
point. For the $\B$ hadron reconstruction, tracks and electromagnetic
calorimeter clusters with no associated track are combined into jets
using a cone algorithm\footnote{Studies have shown that the cone
  jet-finder provides the best $\B$ hadron energy and direction
  resolution compared to other jet
  finders~\cite{bib-opalbss}.}~\cite{bib-jetcone} with a cone
half-angle of $0.65\rm\,rad$ and a minimum jet energy of $5.0\GeV$.
The two most energetic jets of each event are assumed to contain the
$\B$ hadrons.  Only jets where the opposite hemisphere yields a
$\qb$-tagging discriminant of at least 0.8, corresponding to a $\qb$
probability of about 80\%, are used in the analysis. The distribution
of the $\qb$-tagging discriminant is shown in Figure~\ref{fig-lbhemi}.

Each remaining jet is searched for secondary vertices using a vertex
reconstruction algorithm similar to that described
in~\cite{bib-bdstar}, making use of the tracking information in both
the $\rphi$ and $\rz$ planes where available. If a secondary vertex is
found, the primary vertex is re-fitted excluding the tracks assigned
to the secondary vertex.  Secondary vertex candidates are accepted and
called `good' secondary vertices if they contain at least three
tracks. If there is more than one good secondary vertex attached to a
jet, the vertex with the largest number of significant\footnote{A
  track is called significant if its impact parameter significance
  with respect to the primary vertex is larger than 2.5. The impact
  parameter significance is defined as the impact parameter of a track
  divided by the uncertainty on this quantity.}  tracks is taken. If
there are two or more such vertices, the secondary vertex with the
larger separation significance with respect to the primary vertex is
taken. Jets without an associated secondary vertex are rejected. This
increases the $\qb$ jet purity and improves the energy resolution of
the $\B$ hadron reconstruction described in the following.

\begin{figure}
  \begin{center}
    \begin{picture}(0.01,0.01)
      \put(-0.2,9){a)}
    \end{picture}
    \epsfig{file=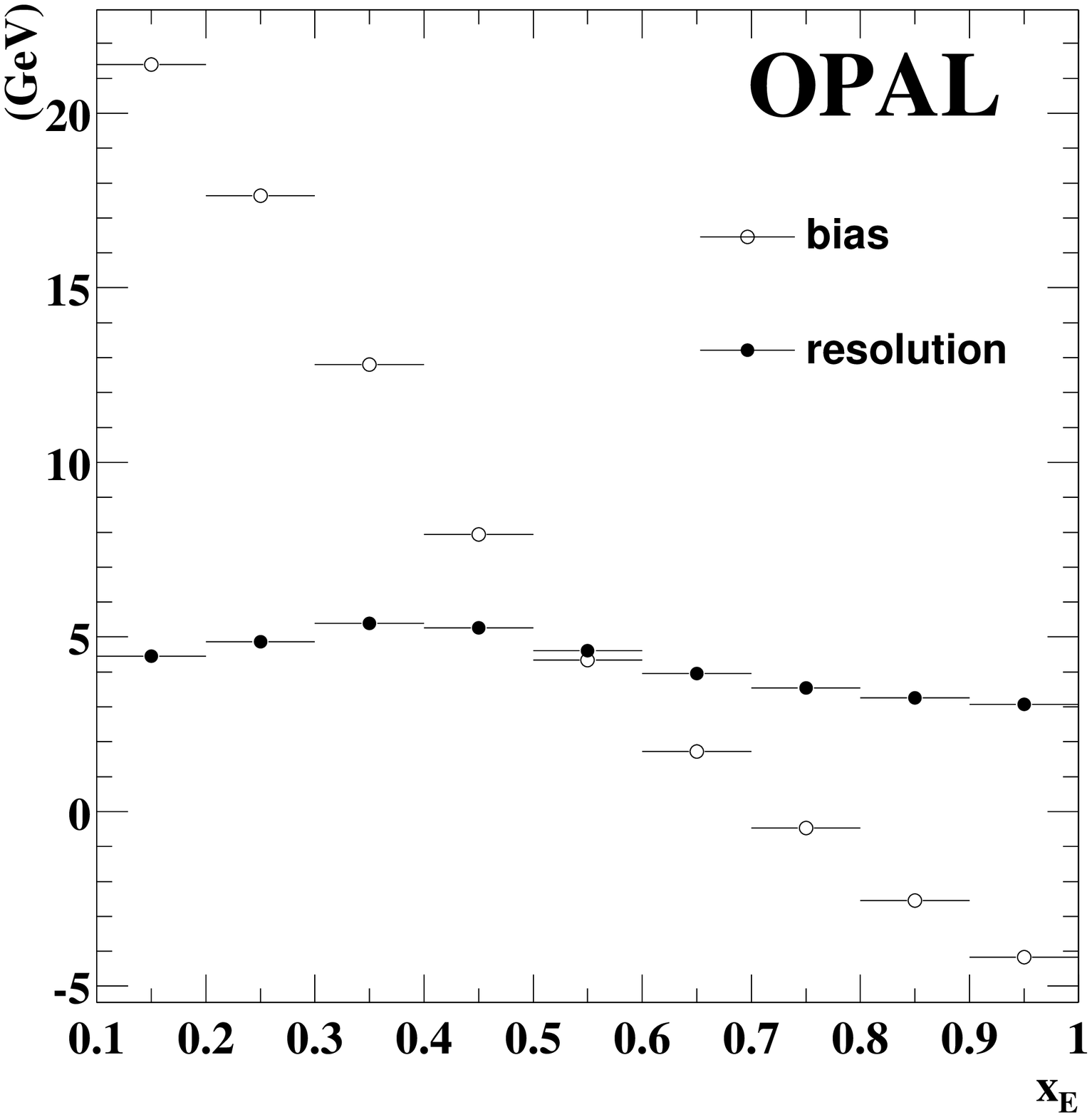,width=10cm}\\[5mm]
    \begin{picture}(0.01,0.01)
      \put(-0.2,9){b)}
    \end{picture}
    \epsfig{file=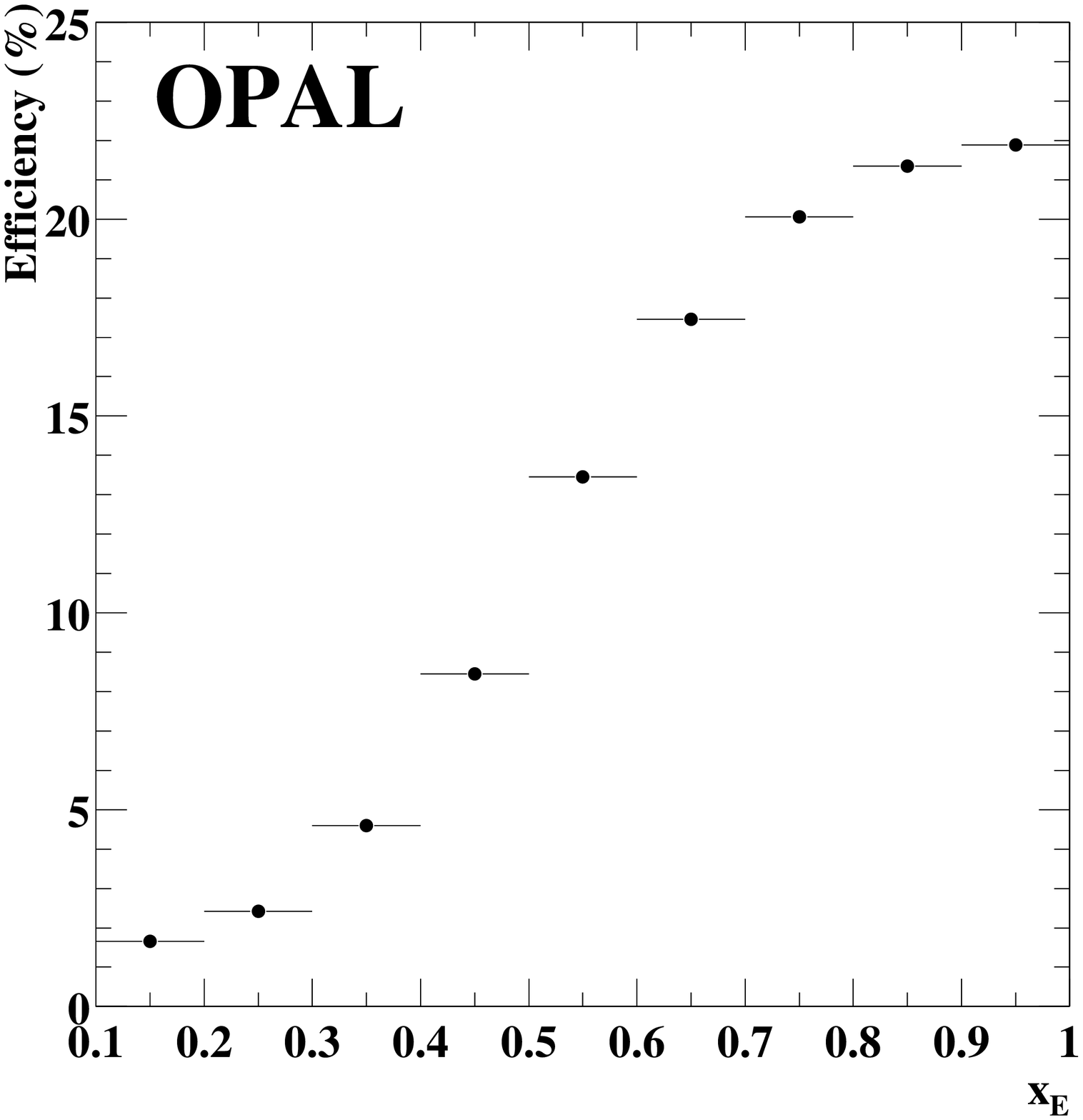,width=10cm}
  \end{center}
  \smcap{\label{fig-eres} a) Dependence of the $\B$ hadron energy
    resolution (black circles) and reconstruction bias (open circles)
    on the generated $\B$ hadron energy. b) Dependence of the $\B$
    hadron reconstruction efficiency on the generated $\B$ hadron energy.}
\end{figure}

Weakly decaying $\B$ hadrons are reconstructed inclusively with a
method described in an earlier publication~\cite{bib-opalbss}. In each
hemisphere defined by the positive axis of the jet found by the cone
algorithm, a weight is assigned to each track and each cluster, where
the weight corresponds to the probability that this track or cluster
is a product of the $\B$ hadron decay. This weight is obtained from
artificial neural networks~\cite{bib-jetnet} exploiting information
from track impact parameters with respect to the primary and secondary
vertices, and from kinematic quantities like the transverse momentum
associated with a track or cluster, measured with respect to the cone
jet axis. A list of all variables is shown in Table~\ref{tab-nnvars}.
The $\B$ hadron momentum is then reconstructed by summing the weighted
momenta of the tracks and clusters. A beam energy constraint assuming
a two-body decay of the $\Z$ and the world average $\B$ meson mass of
$5.279\GeVcc$~\cite{bib-pdg} for the $\B$ hadron is applied to improve
the energy resolution.  The constraints lead to a biased energy
reconstruction, particularly when the true $\B$ hadron energy is very
small, as can be seen in Figure~\ref{fig-eres}a. However, only a small
fraction of the data sample is in the low-energy region affected by a
large bias. For most events, in the peak of the $\B$ hadron energy
distribution, the bias is small, and all biases are taken into account
by the fitting procedures used in both the model-dependent and
model-independent analyses. Possible systematic uncertainties
arising from the biased energy reconstruction are discussed in
Section~\ref{sec-systematics} of this paper. The energy of the weakly
decaying $\B$ hadron is expressed in terms of the scaled energy
$\xE=E_\B / E_{beam}$, where $E_{beam}=\sqrt{s}/2$ is the LEP beam
energy for the event. The quantity $\xE$ is restricted to values above
$5.279\GeV/E_{beam}\approx0.1$ by the $\B$ meson mass constraint, and
it cannot exceed 1.0 due to the beam energy constraint.

\begin{table}
\begin{center}
\begin{tabular}{|l|}
\hline
track neural network \\
\hline
\hline
track momentum \\
track rapidity with respect to estimated $\B$ hadron flight direction \\
track impact parameter with respect to primary vertex in $\rphi$
    projection ($d_0$) \\
track impact parameter with respect to primary vertex in $z$ projection
    ($z_0$) \\
$d_0$ impact parameter significance \\
$z_0$ impact parameter significance \\
3d impact parameter significance with respect to the primary vertex \\
3d impact parameter significance with respect to the secondary vertex \\
\hline
\hline
cluster neural network \\
\hline
\hline
cluster energy \\
cluster rapidity with respect to estimated $\B$ hadron flight direction \\
\hline
\end{tabular}
\end{center}
\smcap{\label{tab-nnvars} Variables used in artificial neural networks
to estimate the probability that a track or calorimeter cluster
originates from a $\B$ hadron decay. The impact parameter significance is
defined as the impact parameter divided by its uncertainty.}
\end{table}

After all these requirements, the distribution of the difference
between the reconstructed energy and that of generated $\B$ hadrons in
simulated data has a rms width of $4.8\GeV$. The energy dependence of
the $\B$ hadron energy resolution is shown in Figure~\ref{fig-eres}a.
The complete $\B$ hadron selection applied to the full data sample
results in $270\,707$ tagged jets with a $\qb$ purity of 96\%. The
average $\B$ hadron selection efficiency is 16\%, with an energy
dependence as shown in Figure~\ref{fig-eres}b.  The measured $\B$
hadron energy distribution, scaled to the beam energy, is shown in
Figures~\ref{fig-fitresult1}--\ref{fig-fitresult4}, and compared to
the various models described in the next section.

\section{Test of hadronisation models}
\label{sec-rewfit}

\begin{table}
  \begin{center}
    \begin{tabular}{|l|c|c|}
      \hline
      fragmentation function & functional form & parameters \\
      \hline
      \hline
      Kartvelishvili et al.~\cite{bib-kartvelishvili} &
      $Nz^{\alpha_\qb}(1-z)$ &
      $\alpha_\qb$ \\
      Bowler~\cite{bib-bowler}\raisebox{-2ex}{\ } &
      $N {1 \over z^{1+bm_\perp^2}} (1-z)^a \exp(-{b m_\perp^2 \over z})$ &
      $a$, $bm_\perp^2$ \\
      Lund symmetric~\cite{bib-lundsymm} &
      $N {1 \over z} (1-z)^a \exp(-{b m_\perp^2 \over z})$ &
      $a$, $bm_\perp^2$ \\
      Peterson et al.~\cite{bib-peterson} &
      $N {1 \over z}(1-{1\over z}-{\varepsilon_\qb \over 1-z})^{-2}$ &
      $\varepsilon_\qb$ \\
      Collins-Spiller~\cite{bib-colspi} &
      $N({1-z \over z} + {(2-z)\varepsilon_\qb \over 1-z})
                  (1+z^2)(1-{1 \over z}-{\varepsilon_\qb \over 1-z})^{-2}$ &
      $\varepsilon_\qb$ \\
      \hline
    \end{tabular}
  \end{center}
  \smcap{\label{tab-fragpar} Fragmentation functions for the JETSET 7.4
     string scheme that are fitted to data in this paper. $N$ is a
     normalisation constant, different for each fragmentation function.}
\end{table}

The $\B$ hadron energy distributions predicted by the JETSET 7.4,
HERWIG 5.9, and HERWIG 6.2 Monte Carlo models are compared to the OPAL
data.  All Monte Carlo simulated events are passed through a detailed
detector simulation~\cite{bib-gopal}. The comparison is performed
using the distribution of the reconstructed scaled energy of the
weakly decaying $\B$ hadron $\xE$.

The HERWIG Monte Carlo uses a parton shower fragmentation followed by
cluster hadronisation model with few parameters. No parameters are
varied in this analysis.  This simplifies the model test to a mere
comparison of the $\xE$ distributions obtained with data and Monte
Carlo simulation. Both HERWIG versions are set up to conserve the
initial $\qb$ quark direction in the $\B$ hadron creation during
cluster decay ({\tt cldir=1}). The main difference between the two
HERWIG samples used in this analysis is that Gaussian smearing of the
$\B$ hadron direction around the initial $\qb$ quark flight direction
is applied in the HERWIG 5.9 sample ({\tt clsmr=0.35}), while smearing
is not used in the HERWIG 6.2 sample ({\tt clsmr(2)=0}).

The JETSET Monte Carlo is based on a parton shower fragmentation
followed by string hadronisation scheme. It requires a fragmentation
function to describe the distribution of the fraction $z$ of the
string light cone momentum that is assigned to a hadron produced at
the end of the string. The JETSET sample in this analysis is
reweighted to use the fragmentation functions of Kartvelishvili et
al.~\cite{bib-kartvelishvili}, Bowler~\cite{bib-bowler}, the Lund
symmetric model~\cite{bib-lundsymm}, and the fragmentation functions
of Peterson et al.~\cite{bib-peterson}, and
Collins-Spiller~\cite{bib-colspi}. The Lund symmetric and Bowler
functions are simplified by assuming the transverse mass of the $\qb$
quark, $m_\perp$, to be constant, which is justified by the smallness
of the average transverse momentum compared to the $\qb$ quark mass. A
further simplification in the Bowler parametrisation is the assumption
of an equality of $\qb$ quark and hadron masses.  The functional forms
of the fragmentation functions are given in Table~\ref{tab-fragpar}.
The parameters of the respective fragmentation functions are fitted to
obtain a best match of the observed $\xE$ distributions in data and
Monte Carlo simulation. In the case of the Peterson et al.,
Collins-Spiller, and Kartvelishvili et al.~models, one free parameter
is available. The Lund and Bowler models each have two free fit
parameters. A $\chi^2$ fit is performed in 46 bins in the $\xE$ range
of 0.5 to 0.95, where in all samples the number of candidates in each
bin is large enough to justify the assumption of Gaussian errors on
the bin content. The fragmentation function and its parameters are
adjusted during the fit by reweighting the Monte Carlo simulated
events, similar to the procedure applied in~\cite{bib-rb}.

\begin{figure}
\begin{center}
\epsfig{file=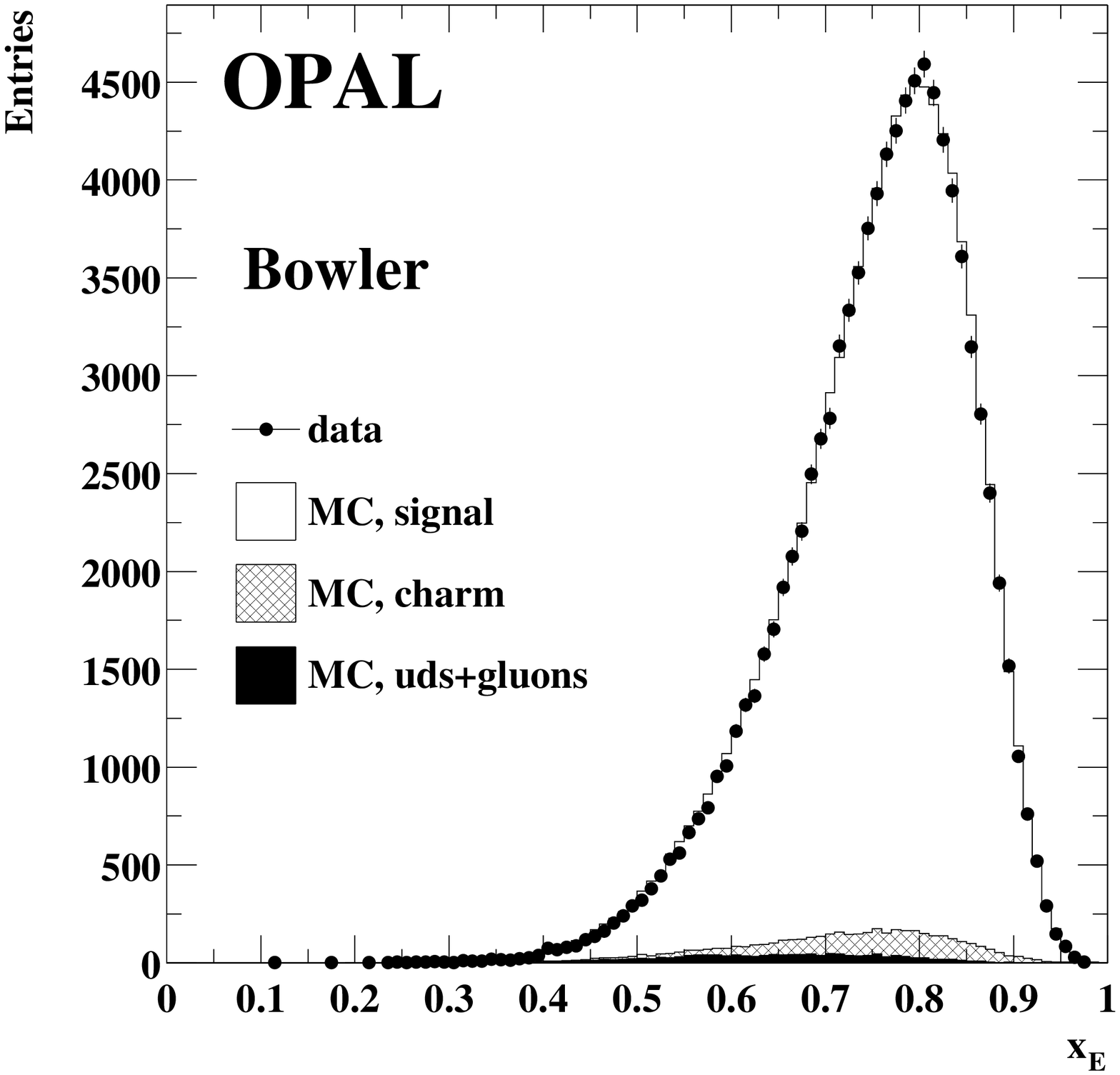,width=10cm}\\[5mm]
\epsfig{file=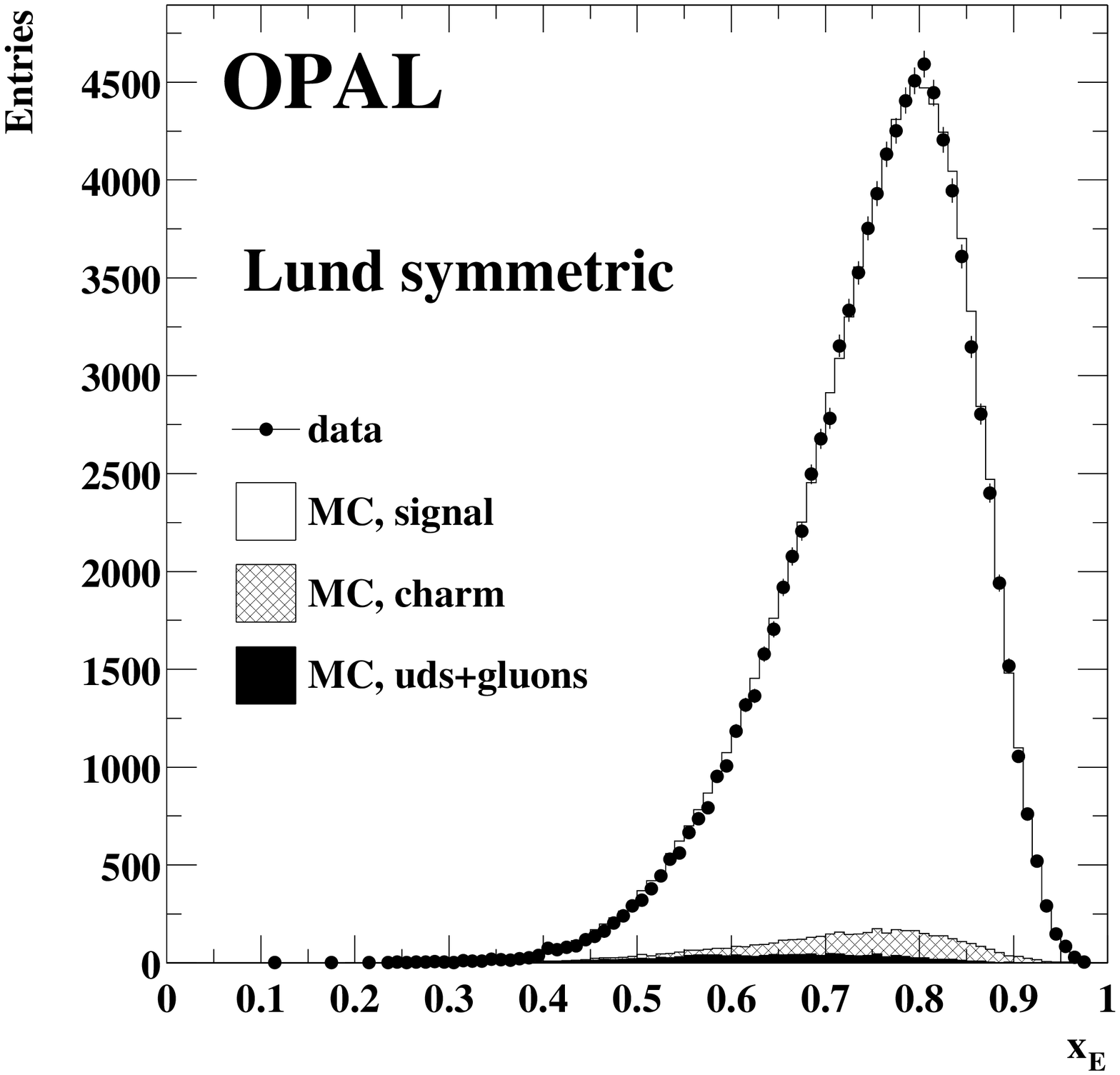,width=10cm}
\end{center}
\smcap{\label{fig-fitresult1} Results of the fit to the data of
  various hadronisation models for JETSET 7.4. The points with error
  bars are the uncorrected reconstructed scaled energy distribution in
  the 1994 data sample.  Only statistical errors are shown.  The
  histogram represents the best match as obtained from the respective
  fragmentation function fits. Background from charm jets is shown as
  hatched histogram, and light quark and gluon background is indicated
  by the black area. Charm jets are preferentially passing the selection
  if the $\qc$ quark flight length is large due to a large boost. The mean
  energy of reconstructed charm candidates is therefore close to that of
  $\qb$ jets.}
\end{figure}

\begin{figure}
\begin{center}
\epsfig{file=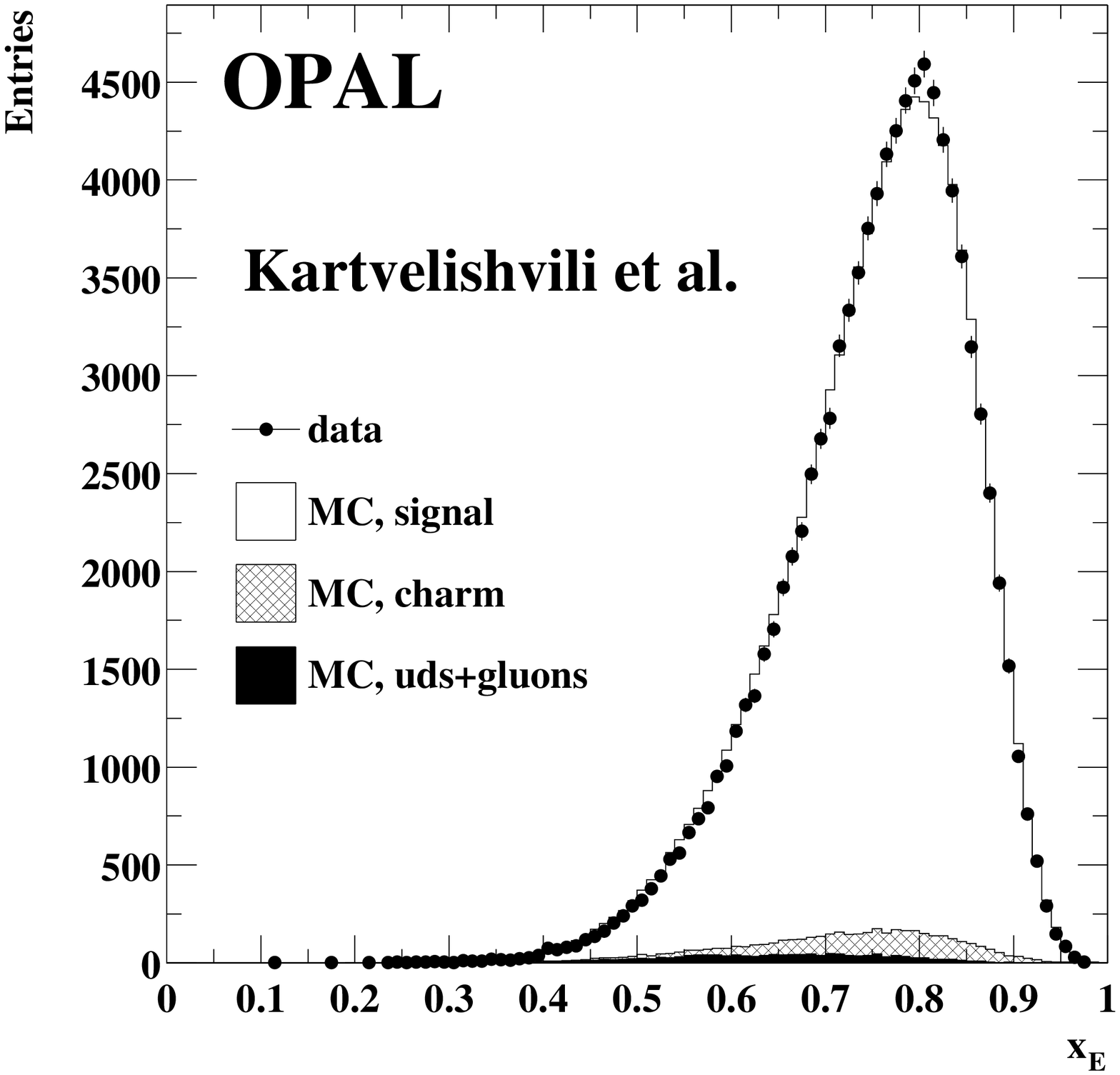,width=10cm}\\[5mm]
\epsfig{file=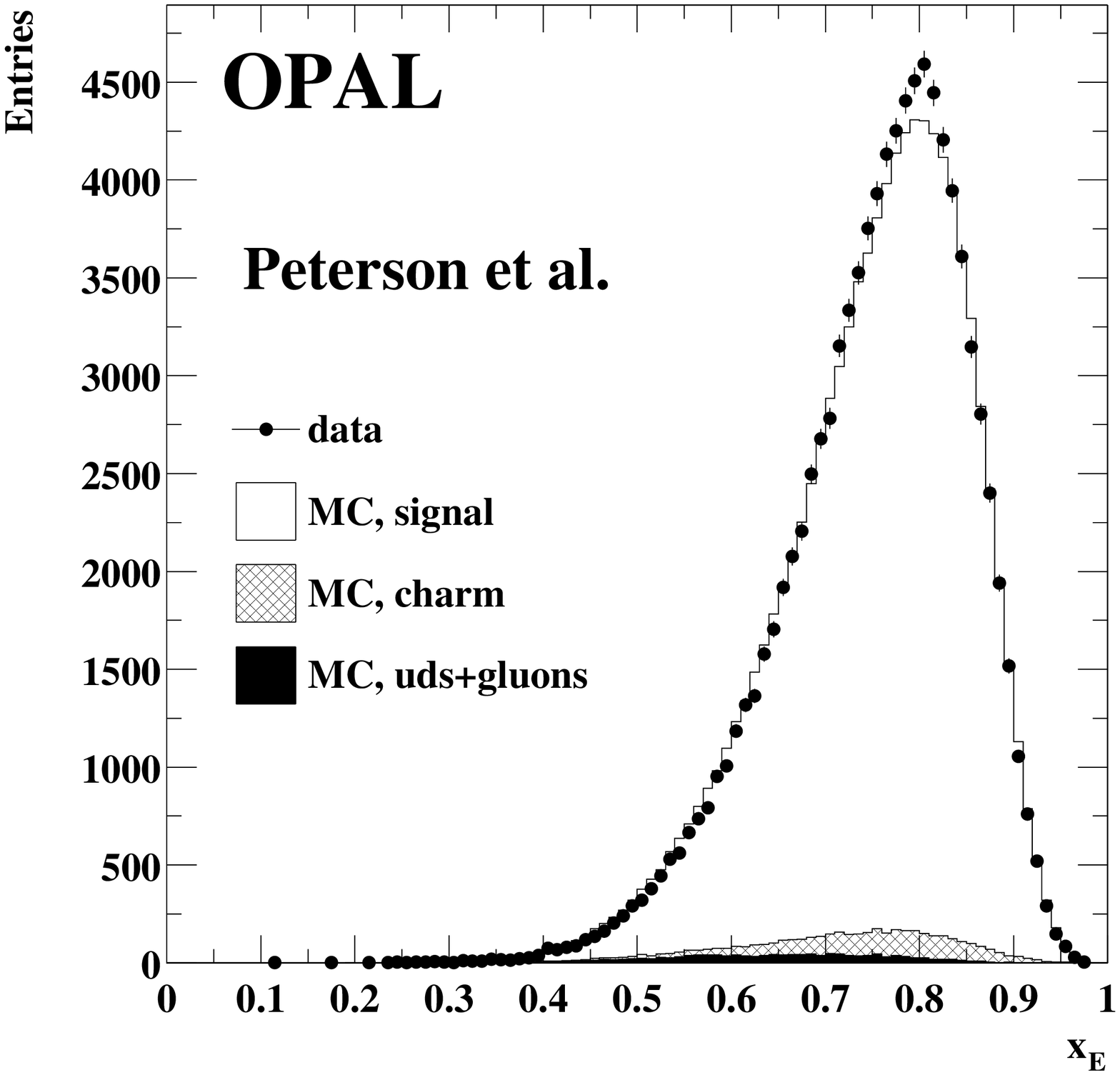,width=10cm}
\end{center}
\smcap{\label{fig-fitresult2} Results of the fit to the data of
  various hadronisation models for JETSET 7.4. The points with error
  bars are the uncorrected reconstructed scaled energy distribution in the 1994
  data sample.  Only statistical errors are shown.  The histogram
  represents the best match as obtained from the respective
  fragmentation function fits. Background from charm jets is shown as
  hatched histogram, and light quark and gluon background is indicated
  by the black area. Charm jets are preferentially passing the selection
  if the $\qc$ quark flight length is large due to a large boost. The mean
  energy of reconstructed charm candidates is therefore close to that of
  $\qb$ jets.}
\end{figure}

\begin{figure}[h]
\begin{center}
\epsfig{file=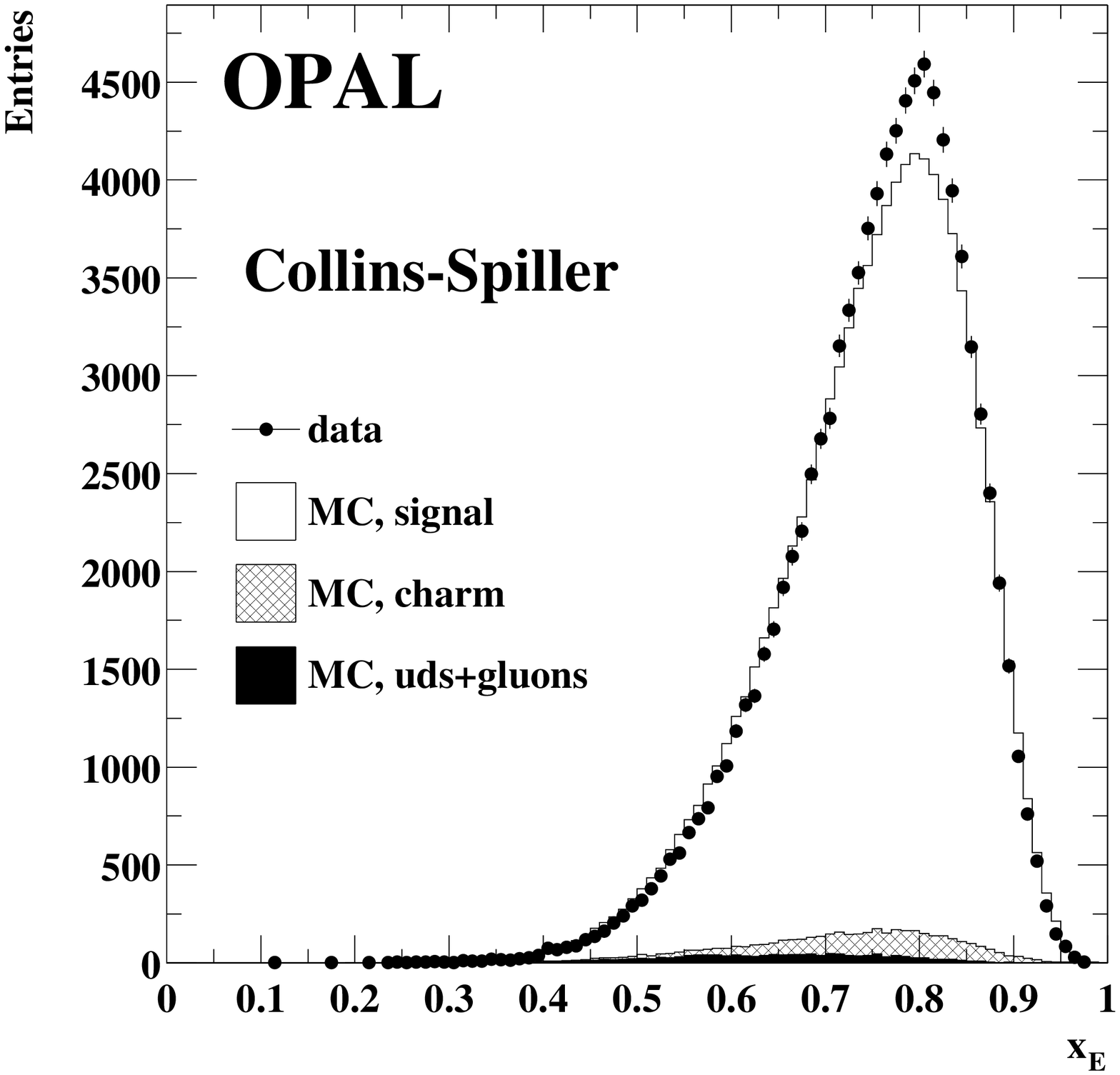,width=10cm}
\end{center}
\smcap{\label{fig-fitresult3} Result of the fit to the data of the
  Collins-Spiller hadronisation model for JETSET 7.4. The points with
  error bars are the uncorrected reconstructed scaled energy distribution in the
  1994 data sample.  Only statistical errors are shown.  The histogram
  represents the best match as obtained from the fragmentation
  function fit. Background from charm jets is shown as
  hatched histogram, and light quark and gluon background is indicated
  by the black area. Charm jets are preferentially passing the selection
  if the $\qc$ quark flight length is large due to a large boost. The mean
  energy of reconstructed charm candidates is therefore close to that of
  $\qb$ jets.}
\end{figure}

\begin{table}[h]
\begin{center}
\begin{tabular}{|l|c|l|r|}
\hline
model\up{6mm}        & parameters          & \multicolumn{1}{|c|}{$\mxE$} & $\chi^2$/d.o.f. \\
\hline
\hline
\dn{Bowler~\cite{bib-bowler}}\up{6mm}
  & $bm_\perp^2=65.1^{+4.8}_{-3.5}\ ^{+16.6}_{-13.9}$
  & \dn{$0.7207^{+0.0008}_{-0.0007}\ ^{+0.0028}_{-0.0029}$}
  & \dn{67/44}  \\
                     & $a=0.80^{+0.08}_{-0.06}\ ^{+0.20}_{-0.21}$ & & \\
\dn{Lund symmetric~\cite{bib-lundsymm}}
  & $bm_\perp^2=15.0^{+1.0}_{-0.7}\pm2.1$
  & \dn{$0.7200^{+0.0009}_{-0.0008}\ ^{+0.0028}_{-0.0030}$}
  & \dn{75/44} \\
                     & $a=1.59^{+0.13}_{-0.10}\pm 0.27$ & & \\
Kartvelishvili et al.~\cite{bib-kartvelishvili}
  & $\alpha_\qb=11.9\pm0.1\pm0.5$
  & $0.7151\pm0.0006\ ^{+0.0020}_{-0.0023}$
  &   99/45 \\
Peterson et al.~\cite{bib-peterson}
  & $\varepsilon_\qb=(41.2\pm0.7\ ^{+3.6}_{-3.5})\times 10^{-4}$
  & $0.7023\pm0.0006\pm0.0019$
  &  159/45 \\
%
% ATTENTION: Peterson result is quoted again in the final discussion!
%            (both epsilon and x_E)
%
Collins-Spiller~\cite{bib-colspi}
  & $\varepsilon_\qb=(22.3^{+0.7}_{-0.6}\ ^{+3.5}_{-4.9})\times 10^{-4}$
  & $0.6870\pm0.0006\ ^{+0.0035}_{-0.0019}$
  &  407/45 \\
HERWIG 6.2           & cldir=1, clsmr(2)=0 & 0.7074 &  540/46 \\
HERWIG 5.9           & cldir=1, clsmr=0.35 & 0.6546 & 4279/46 \\
\hline
\end{tabular}
\end{center}
\smcap{\label{tab-fitresult} Results of the comparison of hadronisation
         models to OPAL data. The parameter fit results and corresponding $\xE$
         values are weighted averages over all datasets from the
         years 1992--2000, where the weights are chosen according to the
         subsample size. The first errors on the parameters are statistical,
         the second systematic. The correlation of the statistical errors
         of $a$ and $bm_\perp^2$ is 98.5\% for the Lund symmetric model, and
         96.4\% for the Bowler fragmentation function.
         The errors on $\mxE$ are the propagated
         statistical parameter errors. The $\chi^2$/d.o.f. values
         are quoted for the 1994 dataset only, which is the largest
         sample. Only statistical errors are included. The errors of the two
         parameters of the Lund and Bowler models
         are almost fully correlated. The parameters given for the HERWIG
         Monte Carlo are not obtained from a fit, but are the values
         used for the generation of each sample.}
\end{table}

\begin{figure}
\begin{center}
\epsfig{file=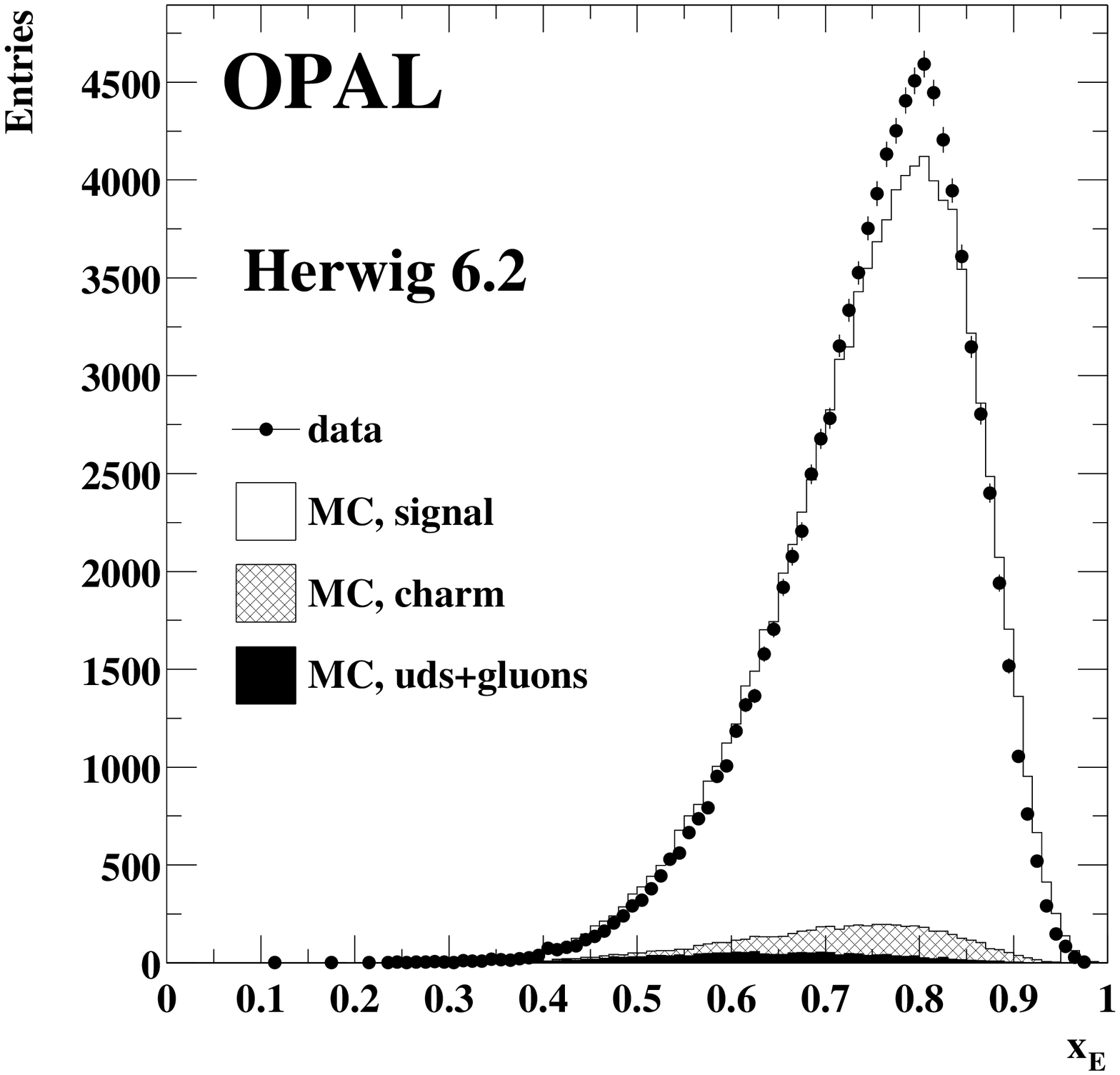,width=10cm}\\[5mm]
\epsfig{file=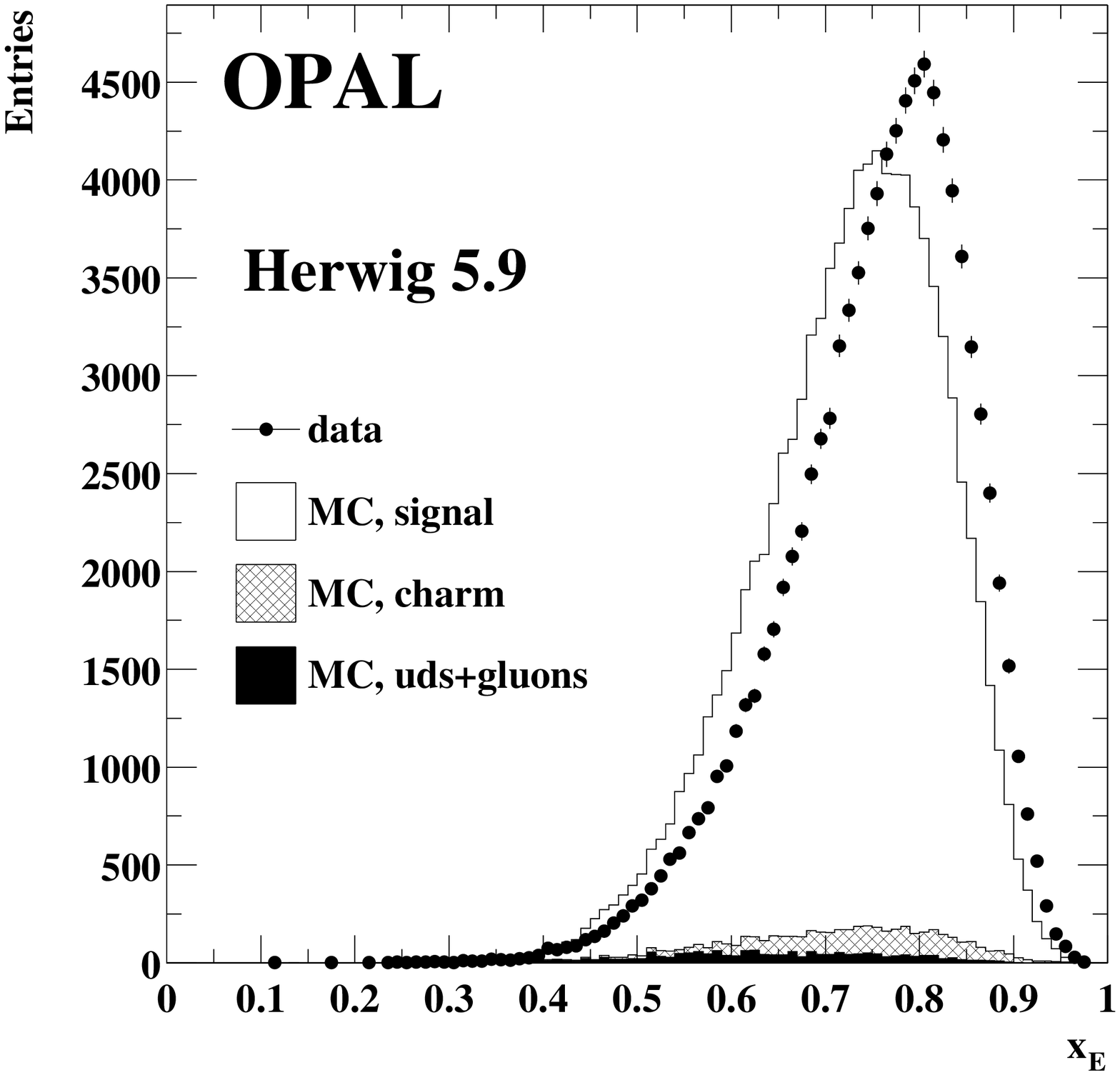,width=10cm}
\end{center}
\smcap{\label{fig-fitresult4} Comparison of different setups of the
  HERWIG Monte Carlo generator with data.  The points with error bars
  are the uncorrected reconstructed scaled energy distribution in the
  1994 data sample.  Only statistical errors are shown.  The histogram
  represents the HERWIG prediction. Background from charm jets is
  shown as hatched histogram, and light quark and gluon background is
  indicated by the black area.}
\end{figure}

The reweighting fit is performed separately for each data sample, and
the fit results are averaged with weights according to the size of the
respective datasets. Consistent results are obtained for all datasets.
The average parameter values are summarised in
Table~\ref{tab-fitresult}. For each parametrisation the corresponding
model-dependent mean scaled energy of weakly decaying $\B$ hadrons is
given. Data samples at $\sqrt{s}=m_\Z + 1.8\GeV$ are excluded in the
calculation of the average $\mxE$.  Table~\ref{tab-fitresult} also
gives a comparison of the fit quality of all JETSET 7.4 fits on the
1994 data, and the HERWIG 5.9 and HERWIG 6.2 results. The fit results
on the 1994 data are shown in
Figures~\ref{fig-fitresult1}--\ref{fig-fitresult4}.  The ordering of
the models according to the goodness of the fits in 1994 data agrees
with all other large data samples; only in a few smaller samples is a
slightly different ordering observed. The quoted $\chi^2$/d.o.f.
values only take into account the statistical uncertainty of data and
Monte Carlo simulation.  Systematic uncertainties are discussed later.
The Bowler, Lund symmetric, and Kartvelishvili et al.~models are
preferred by the data, with respective $\chi^2$/d.o.f. values of
67/44, 75/44, and 99/45 in the 1994 sample.
Figures~\ref{fig-fitresult1}--\ref{fig-fitresult4} show that the
Peterson et al.~and Collins-Spiller parametrisations for JETSET, as
well as the HERWIG 6.2 model, are too broad. The HERWIG 5.9 model is
too soft.

\boldmath
\section{Model-independent measurement of $\mxE$}
\unboldmath

In the previous section, information was extracted from the observed
energy distribution making explicit use of a set of models to describe
the data. In this section, a measurement of the mean scaled energy of
$\B$ hadrons, $\mxE$, outside a specific model framework will be
presented. This is accomplished by unfolding the observed energy
distribution.

Two complementary unfolding procedures are used to obtain an estimate
of the true $\xE$ distribution from the observed distribution of the
reconstructed scaled $\B$ hadron energy. In both cases the amount and
energy distribution of background in the $\B$ hadron candidate sample
is estimated from the Monte Carlo simulation and subtracted from the
data.

The main method starts by fitting the observed data $\xE$
distribution, and the observed and the true $\xE$ distribution in the
Monte Carlo simulation with smooth functions (splines). The true and
observed Monte Carlo distributions are then reweighted simultaneously
until the observed $\xE$ distribution agrees in data and simulation.
The reweighted true $\xE$ distribution of the Monte Carlo simulation
then provides an estimate of the corresponding distribution in data.
Details of how the result is stabilised are described later. This
method is almost independent of the initial Monte Carlo distribution
and thus reduces model-dependence in the unfolding process.
Furthermore, the result is represented as an unbinned spline function,
which is optimal for the calculation of the mean value of the unfolded
distribution.  This algorithm is coded using the software package
RUN~\cite{bib-unfoldRUN} and was already used
in~\cite{bib-earlierRUN}.

The second approach makes use of the SVD-GURU software
package~\cite{bib-unfoldGURU}. The correspondence between the observed
and true $\B$ hadron energy distributions in the Monte Carlo
simulation is represented by a $20\times20$ matrix. The unfolding
process comprises a matrix inversion to obtain an estimate of the true
data $\xE$ distribution from the observed distribution.  In this
approach, the model dependence was found to be stronger than when
using the RUN program. Furthermore, a coarse binning appropriately
adapted to the detector's resolution and the amount of available data
might lead to systematic effects when describing the energy
distribution in terms of its mean value. Therefore SVD-GURU is only
used to cross-check the result obtained by RUN and to provide an
estimate of the systematic uncertainty due to unfolding.

\begin{figure}
\begin{center}
 \epsfig{file=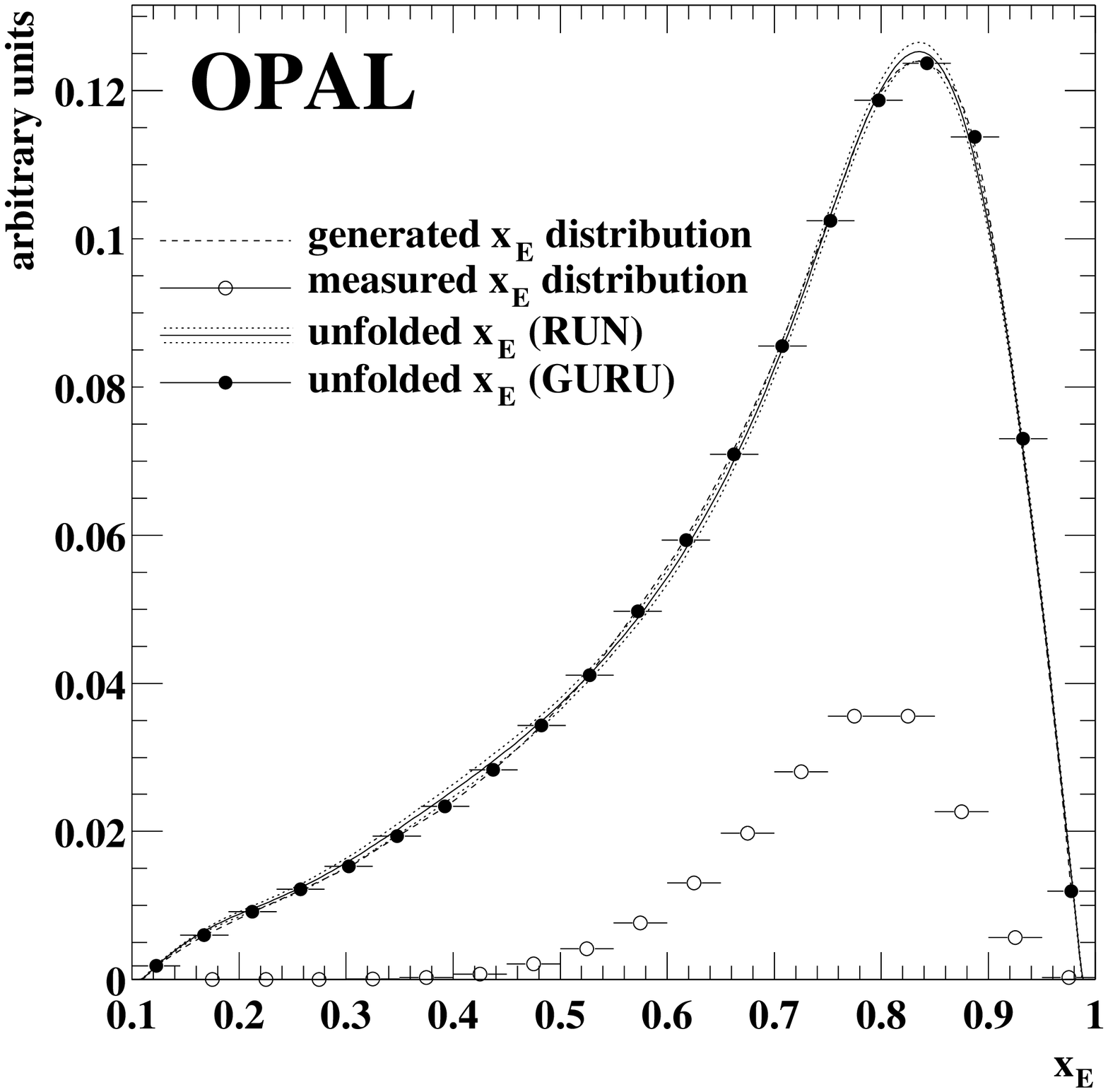,width=10cm}
\end{center}
\smcap{\label{fig-unfperf} Performance of the unfolding algorithms
  used in this analysis. The dashed line represents the generated
  scaled energy distribution of weakly decaying $\B$ hadrons. Open
  circles with error bars represent the observed $\xE$ distribution
  for Monte Carlo simulated events corresponding to the 1994 OPAL
  detector setup. Shape and normalisation are different from the
  generated distribution due to limited and energy-dependent
  efficiency, detector resolution and reconstruction bias. Full
  circles with error bars and the solid line with error band indicate
  the SVD-GURU and RUN unfolding results for this sample.}
\end{figure}

Raw unfolding solutions often oscillate strongly around the correct
solution. In the case of a binned representation of the data this
effect can simply be understood by strong negative bin-by-bin
correlations introduced by the finite detector resolution.  Both
methods used here suppress these oscillations by limiting the number
of degrees of freedom of the unfolding solution.  The RUN algorithm
represents the unfolding result as expansion into a set of orthogonal
functions. The uncertainties on the coefficients of these functions
are determined, and only those functions with coefficients
significantly different from zero are taken into account. SVD-GURU
rotates the unfolding matrix to estimate its effective rank. The
unfolding is then performed in a rotated space with a smaller matrix
including only the significant contributions. The number of degrees of
freedom used for the unfolding procedure was found to agree in the RUN
and SVD-GURU approaches in all subsamples described below. A further
means of regularisation is available in the RUN package. Of all
remaining solutions to the unfolding problem, one is chosen that
minimises the integral over the squared first derivative of the
unfolding solution. Monte Carlo studies show that this regularisation
leads to essentially bias-free results on all samples. The performance
of the unfolding algorithms is illustrated in
Figure~\ref{fig-unfperf}.

\begin{figure}
\begin{center}
 \epsfig{file=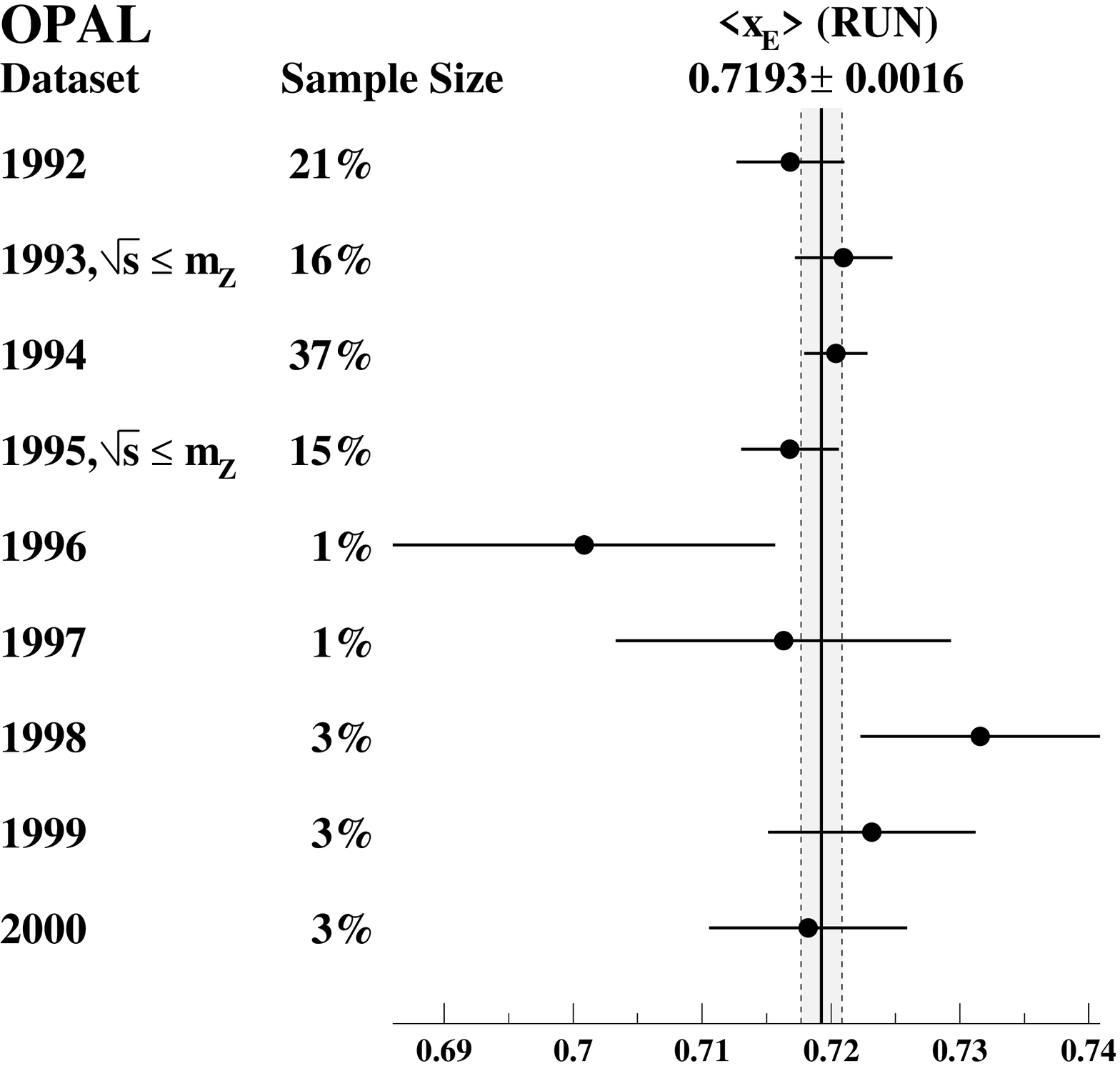,width=10cm}
\end{center}
\smcap{\label{fig-unfspread} Unfolding results of all data samples
  with a centre-of-mass energy on or below the $\Z$ mass, obtained
  with the RUN unfolding package. The fraction of the total data
  sample contributed by each subsample is given on the left.  The
  vertical line indicates the weighted average $\mxE$, and the shaded
  region represents the uncertainty of this average.}
\end{figure}

The unfolding is performed separately for data from all years of
1992--2000. The 1993 and 1995 datasets at a centre-of-mass energy
below $m_{\Z}$ show $\xE$ distributions that are compatible with those
taken at the $\Z$ resonance peak in the Monte Carlo simulation. The
1993 and 1995 datasets at $m_{\Z}+1.8\GeV$ show a significantly lower
$\mxE$, caused by a large amount of initial state radiation at this
energy. In this case the quark energy prior to fragmentation is lower
on average than the beam energy. As the beam energy is used as an
estimator of the quark energy prior to fragmentation, the average
$\xE$ value is lower than in samples without significant initial state
radiation. The $m_\Z+1.8\GeV$ samples are therefore analysed
separately.

Both RUN and SVD-GURU analyses are performed with a Monte Carlo
simulation that is reweighted to match the best result of the
model-dependent reweighting fits for the respective datasets. This
procedure is also followed by SLD in their latest $\qb$ hadronisation
analysis~\cite{bib-bfrsld02}. The goal is to reduce the dependence of
the unfolding result on the Monte Carlo sample used for unfolding. The
effect of not using the best parametrisation, but the second and third
best instead, are studied below as a systematic effect.  JETSET 7.4
Monte Carlo simulation samples are used to obtain the central result,
and will be compared to unfolding results using HERWIG in the
discussion of systematic effects below.

\begin{figure}
\begin{center}
 \epsfig{file=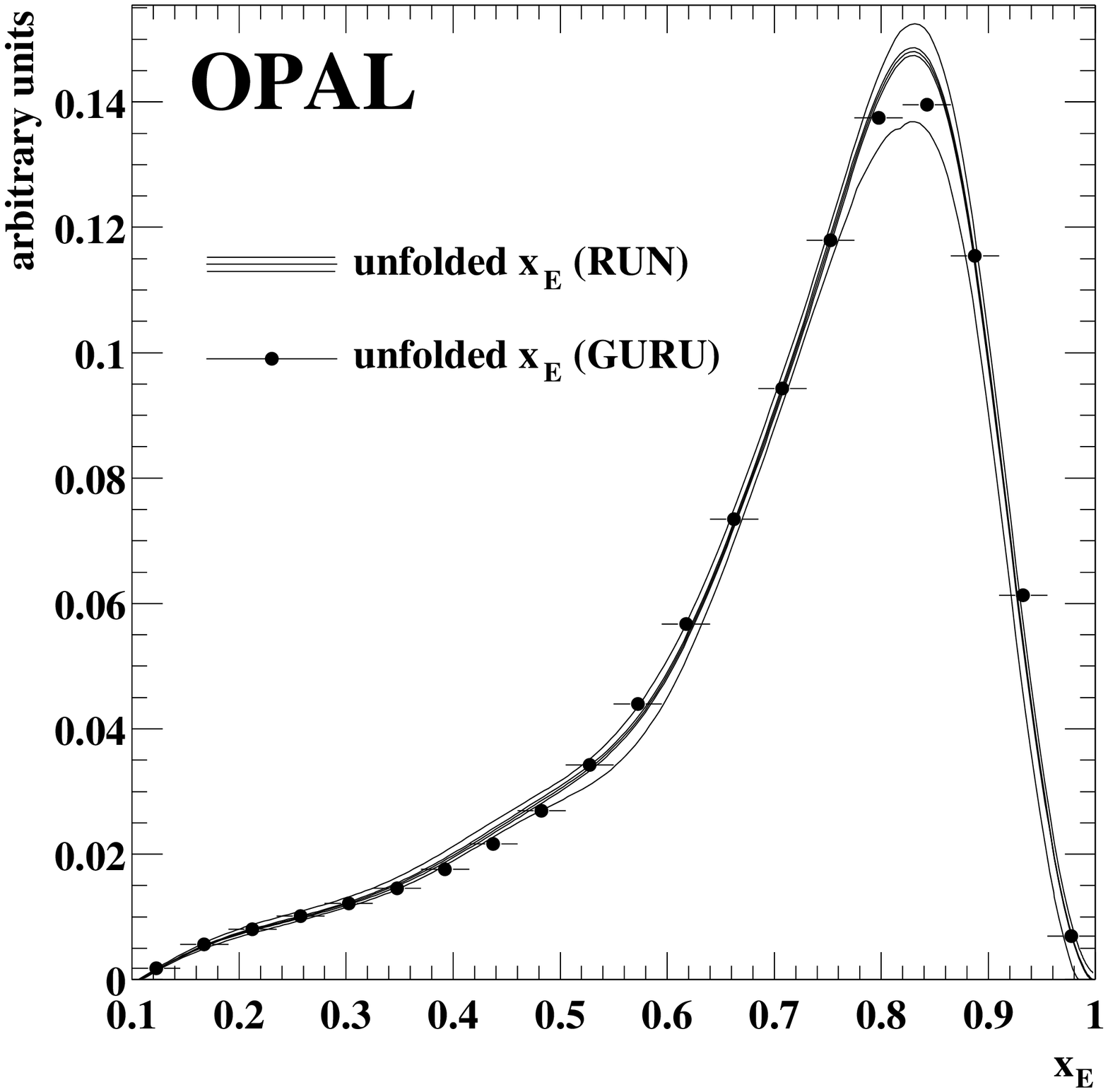,width=10cm}
\end{center}
\smcap{\label{fig-unfresult} Spline representation of the RUN
  unfolding result (line with error band), and binned GURU unfolding
  result (points with error bars), for the full data sample at
  $\sqrt{s}\leq m_\Z$. The narrower error band around the RUN
  unfolding results corresponds to the statistical uncertainty, the
  broader error band represents the total uncertainty.}
\end{figure}

The unfolding result for all data samples at or below the $\Z$
resonance is shown in Figure~\ref{fig-unfspread}. The mean scaled $\B$
hadron energy, obtained with the RUN unfolding algorithm and averaged
over all datasets by using the subsample size for the weights, is
\[
  \mxE = 0.7193 \pm 0.0016(stat) \ ,
\]
where the uncertainty includes the statistical uncertainties due to
limited data and Monte Carlo sample sizes, and the statistical
uncertainty on the Monte Carlo efficiency.  Consistent $\mxE$ values
were obtained for the individual data samples (see
Figure~\ref{fig-unfspread}).  This model-independent measurement
agrees with the $\mxE$ values in the framework of the best models as
seen in Table~\ref{tab-fitresult}.  The unfolding result spline for
the full dataset is plotted in Figure~\ref{fig-unfresult}.

The mean scaled energy observed in the $m_{\Z}+1.8\GeV$ samples is
found to be
\[
  \mxE = 0.7130 \pm 0.0056(stat) \ .
\]
The 1993 and 1995 $m_{\Z}+1.8\GeV$ data samples give consistent values
of $\mxE$.  The difference of the results for the different energies
is consistent with the prediction obtained from Monte Carlo samples at
similar energies.

The results obtained with the unfolding program SVD-GURU ($0.7195\pm
0.0015(stat)$ for the main dataset, $0.7152\pm 0.0053(stat)$ for the
$m_{\Z}+1.8\GeV$ samples) are in very good agreement with the ones
achieved with the RUN package.  This is also demonstrated in
Figure~\ref{fig-unfresult}, where the results obtained from the full
$\sqrt{s}\leq m_\Z$ data sample by both algorithms are compared. The
statistical uncertainties of both methods are also very similar.

\section{Systematic uncertainties}
\label{sec-systematics}

Given the large data sample collected with the OPAL detector, and the
inclusive character of the analysis presented here, the statistical
uncertainties on the results of the previous sections are expected to
be small compared to the systematic uncertainties introduced by
limited knowledge of physics parameters which possibly affect the
measured quantities. In this section, an overview of all systematic
checks is given for the reweighting fit and for the unfolding
analysis.

The distribution of the $\qb$-tagging discriminant in data agrees well
with the Monte Carlo simulation. However, it is necessary for this
analysis to ensure that this is separately true for different $\B$
hadron energy regions. Therefore the $\qb$-tagging discriminant was
investigated in ten bins in $\xE$. The ratio of the $\qb$-tagging
discriminant distributions of data and Monte Carlo simulation is
calculated and fitted by a linear function in each energy bin
separately. The slope of this function is an indicator of the quality
of agreement of data and simulation. All fitted slopes are compatible
with zero within two standard deviations.

\subsubsection*{Systematic uncertainties for fits to various models}

The systematic checks performed for the fits to models are listed
below.  The resulting systematic uncertainty estimates for the mean
scaled energy $\mxE$ in the framework of the respective models are
summarised in Table~\ref{tab-fitsysterr}.
\begin{itemize}
\item The energy resolution of the OPAL calorimeters in the Monte
  Carlo simulation is varied by $\pm10\%$ relative to its central
  value. This range is motivated by jet energy resolution studies in
  two-jet events, where a difference between the resolution in data
  and the Monte Carlo simulation of 3.6\% was found in some datasets.
  Under the assumptions that 50\% of the total jet energy are
  contributed by neutral particles and that the observed difference
  can be fully accounted to the calorimeters, the above variation
  range covers this effect.
\item Similar studies indicate a possible difference of the energy
  scale between data and simulation of up to 0.4\% in some datasets.
  The energy scale is varied within this range, and the resulting
  difference of the fit results is taken as systematic uncertainty.
\item The resolution of the $\rphi$-related track parameters $d_0$
  (transverse impact parameter), $\phi_0$ (initial azimuth), and
  $\kappa$ (curvature) is varied in the range
  $\pm10\%$~\cite{bib-rb}.
\item The resolution of the $\rz$-related track parameters $z_0$
  (longitudinal impact parameter) and $\tan{\lambda}$ (dip angle,
  $\lambda=\pi/2-\theta$) is varied by $\pm10\%$~\cite{bib-rb}.
\item Figure~\ref{fig-eres}a shows a large energy reconstruction bias
  for low energy $\B$ hadrons. Both the number of low energy $\B$
  hadrons and the efficiency for reconstructing them are small.  This
  large bias therefore only affects a very small fraction of the
  candidates. The effect of possible mismodeling of the bias in the
  Monte Carlo simulation is evaluated by varying the bias around the
  values that describe the data best. The reconstruction bias in the
  high energy region cannot be changed by more than $\pm1\%$ without
  leading to a significant degradation of the agreement between data
  and Monte Carlo simulation. The bias for low energy $\B$ hadrons,
  with a reconstructed $\xE$ below 0.6, is varied by $\pm10\%$. This
  range is also motivated by significant degradation of the agreement
  between data and Monte Carlo simulation. The largest deviations of
  the measured quantities are taken as systematic uncertainties.
\item The relative fraction of different $\B$ hadron species in the
  sample of primary $\B$ hadrons influences the measurement, because
  different $\B$ hadron species have different energy distributions.
  All values obtained in this analysis are calculated with Monte Carlo
  samples that are reweighted to reflect the current best knowledge of
  the hadron fractions. The associated systematic uncertainty is
  estimated by varying the $\qb$ baryon fraction within the range
  $(10.3\pm1.8)\%$~\cite{bib-bcomb} given by the average of the
  LEP/SLD/CDF measurements of this quantity. The fraction of $\Bs$ in
  the sample is varied in the range of
  $(9.8\pm1.2)\%$~\cite{bib-bcomb}.
\item The amount of orbitally excited $\Bss$ mesons has been measured
  by all LEP collaborations~\cite{bib-bdstar,bib-bssrates}. An error
  weighted average of the LEP measurements is $(28.4\pm3.5)\%$, and
  the fraction of orbitally excited $\Bss$ mesons is varied within
  this range.
\item The Q-value of orbitally excited $\Bss$ mesons~\cite{bib-pdg} is
  about $40\MeV$ smaller than the value in the Monte Carlo samples
  used in this analysis. All results are corrected for this effect,
  and the difference to the values obtained without correction is
  taken as systematic uncertainty.
\item The average multiplicity of charged particles from a $\B$ hadron
  decay was found at LEP to be $4.955\pm0.062$~\cite{bib-bcomb}, and is
  varied within this range.
\item The average lifetime of weakly decaying $\B$ hadrons affects the
  efficiency of secondary vertex reconstruction, and is
  varied in the range $(1.577\pm0.016)\ps$~\cite{bib-pdg}.
\item The average lifetime of weakly decaying charm hadrons determines
  the amount of charm background found in the $\B$ hadron candidate
  sample. The $\Dzero$, $\Dplus$, $\Ds^+$, and $\Lambdac^+$ lifetimes
  are varied within the uncertainties quoted in~\cite{bib-pdg}.
\item The charm background in the Monte Carlo simulation samples is
  reweighted to the same hadronisation model as the $\B$ hadron
  distribution in the respective fits. The central values and
  uncertainties of the charm fragmentation function are taken from
  earlier OPAL measurements~\cite{bib-rcpaper}. For the evaluation of
  the systematic uncertainty the parameters are varied within their
  uncertainties.
\item Jets from gluon splitting to $\bb$ quark pairs are treated as
  background in this analysis. To account for the uncertainty of the
  average LEP result of the gluon splitting rate, $0.00254\pm0.00050$
  $\bb$ pairs per hadronic event~\cite{bib-bcomb}, the rate is
  varied within this range.
\item The number of $\cc$ pairs from gluon splitting per hadronic
  event is varied within the LEP uncertainty of
  $0.0299\pm0.0039$~\cite{bib-bcomb}.
\item The partial width of the $\Z$ into $\bb$ quark pairs, normalised
  to the total hadronic width of the $\Z$, is measured to be
  $\Rb=0.21646\pm0.00065$~\cite{bib-pdg}. Varying this fraction within
  the quoted uncertainty leads to varying background levels in the
  unfolding Monte Carlo sample. This causes negligible changes of the
  fit results.
\item The analogous quantity for charm quark pairs, $\Rc$, is less
  well known, with a current best value of
  $0.1719\pm0.0031$~\cite{bib-pdg}. The impact on the fit results of
  varying $\Rc$ within this range is negligible.
\item Limited knowledge of the LEP beam energy produces a
  corresponding uncertainty on $\xE$, although the dependency of $\xE$
  on the beam energy is reduced due to the fact that the beam energy
  also enters the calculation of the reconstructed $\B$ hadron energy
  via the beam energy constraint. The assumed LEP beam energy is
  varied within $\pm 8\MeV$, which is the largest reported uncertainty
  for any sample at or close to the $\Z$
  resonance~\cite{bib-lepenergy}.  A correlation of 100\% between the
  resulting uncertainties for the different data taking years is
  assumed.
\item The parameter values depend slightly on the $\xE$ range used for
  the fit. The lower end of the fit range is varied within
  $\xE=0.5\pm0.1$, and the upper range is varied within
  $\xE=0.95\pm0.05$. The largest deviation from the result obtained
  using the central values is taken as the systematic uncertainty.
\item The bin width used in the fit is varied by $\pm10\%$. The
  maximum deviation from the result with standard binning is used to
  estimate the associated systematic uncertainty.
\end{itemize}

\begin{table}
\begin{center}
\begin{tabular}{|l|r|r|r|r|r|}
\hline
& \multicolumn{1}{c|}{Kartvelishvili} & \multicolumn{1}{c|}{Bowler} & \multicolumn{1}{c|}{Lund} & \multicolumn{1}{c|}{Peterson} & \multicolumn{1}{c|}{Collins-Spiller} \\
& \multicolumn{1}{c|}{et al.~\cite{bib-kartvelishvili}} & \multicolumn{1}{c|}{\cite{bib-bowler}} & \multicolumn{1}{c|}{\cite{bib-lundsymm}} & \multicolumn{1}{c|}{et al.~\cite{bib-peterson}} & \multicolumn{1}{c|}{\cite{bib-colspi}} \\
\hline
\hline
%               % 
%============================================================================================================
energy resolution  \up{6mm}
                & $\pm0.0010$ 
                & $\pm0.0015$
                & $\pm0.0016$
                & $\pm0.0008$
                & $\pm0.0010$ \\
energy scale    & $\pm0.0003$
                & $\pm0.0004$
                & $^{+0.0004}_{-0.0005}$
                & $\pm0.0003$
                & $\pm0.0003$ \\
$\rphi$ tracking& $\pm0.0015$ 
                & $\pm0.0021$
                & $\pm0.0021$
                & $\pm0.0014$
                & $\pm0.0013$ \\
$z$ tracking    & $\pm0.0002$ 
                & $\pm0.0003$
                & $\pm0.0003$
                & $\pm0.0001$
                & $\pm0.0001$ \\
bias modeling   & $<0.0001$
                & $\pm0.0001$
                & $<0.0001$
                & $\pm0.0001$
                & $\pm0.0003$ \\
$\qb$ baryons   & $\pm0.0002$
                & $^{+0.0001}_{-0.0000}$
                & $<0.0001$
                & $\pm0.0002$
                & $\pm0.0002$ \\
$\Bs$ fraction  & $\pm0.0002$
                & $\pm0.0002$
                & $\pm0.0002$
                & $^{+0.0002}_{-0.0001}$
                & $\pm0.0001$ \\
$\Bss$ fraction & $\pm0.0004$
                & $\pm0.0001$
                & $\pm0.0001$
                & $\pm0.0003$
                & $\pm0.0003$ \\
$\Bss$ Q-value  & $<0.0001$
                & $<0.0001$
                & $<0.0001$
                & $<0.0001$
                & $<0.0001$ \\
$\qb$ decay multiplicity
                & $\pm0.0002$
                & $\pm0.0004$
                & $\pm0.0004$
                & $\pm0.0002$
                & $\pm0.0002$ \\
$\qb$ lifetime  & $\pm0.0001$
                & $\pm0.0001$
                & $\pm0.0001$
                & $\pm0.0001$
                & $\pm0.0001$ \\
$\qc$ lifetime  & $^{+0.0001}_{-0.0002}$
                & $^{+0.0001}_{-0.0003}$
                & $^{+0.0001}_{-0.0002}$
                & $^{+0.0001}_{-0.0002}$
                & $^{+0.0001}_{-0.0002}$ \\
charm fragmentation
                & $\pm0.0001$
                & $\pm0.0001$
                & $\pm0.0001$
                & $\pm0.0001$
                & $^{+0.0000}_{-0.0001}$ \\
$\gbb$ rate     & $<0.0001$
                & $<0.0001$
                & $\pm0.0001$
                & $<0.0001$
                & $<0.0001$ \\
$\gcc$ rate     & $\pm0.0001$
                & $\pm0.0002$
                & $\pm0.0002$
                & $\pm0.0001$
                & $\pm0.0001$ \\
$\Rb$           & $<0.0001$
                & $<0.0001$
                & $<0.0001$
                & $<0.0001$
                & $<0.0001$ \\
$\Rc$           & $<0.0001$
                & $<0.0001$
                & $<0.0001$
                & $<0.0001$
                & $<0.0001$ \\
beam energy     & $\pm0.0001$
                & $\pm0.0001$
                & $\pm0.0001$
                & $\pm0.0001$
                & $\pm0.0001$ \\
fit range       & $^{+0.0004}_{-0.0011}$
                & $^{+0.0004}_{-0.0009}$
                & $^{+0.0005}_{-0.0012}$
                & $^{+0.0009}_{-0.0008}$
                & $^{+0.0030}_{-0.0008}$ \\
binning effects & $\pm0.0002$
                & $\pm0.0003$
                & $^{+0.0003}_{-0.0002}$
                & $<0.0001$
                & $<0.0001$ \\
\hline
total \up{6mm}  & $^{+0.0020}_{-0.0023}$
                & $^{+0.0028}_{-0.0029}$
                & $^{+0.0028}_{-0.0030}$
                & $\pm0.0019$
                & $^{+0.0035}_{-0.0019}$ \\
\hline
\end{tabular}
\end{center}
\smcap{\label{tab-fitsysterr} Overview of systematic uncertainty contributions to the
 model-dependent $\mxE$ measurements.}
\end{table}

The JETSET 7.4 Monte Carlo samples used for the reweighting fit were
generated using the Peterson et al.~fragmentation function with
$\varepsilon_\qb=38\times10^{-4}$. The fit result for the Peterson et
al.~function in data is $\varepsilon_\qb=(41.2\pm0.7)\times 10^{-4}$.
The fact that the Monte Carlo tuning and the data fit result are close
has the advantage that adverse effects due to weights far from 1.0 are
not expected.  However, additional studies were performed to verify
that the closeness of the two values is not introduced by improper
reweighting. An older sample of 4 million hadronic JETSET 7.4 events
with Peterson et al.~fragmentation function with
$\varepsilon_\qb=57\times10^{-4}$ is used to repeat the fit for the
1994 data sample. The fit result obtained with this sample
($\varepsilon_\qb=(40.3\pm1.1)\times10^{-4}$) is in agreement with the
result obtained with the main $\varepsilon_\qb=38\times10^{-4}$ 1994
Monte Carlo sample ($\varepsilon_\qb=(40.6\pm1.0)\times10^{-4}$).

\subsubsection*{Systematic uncertainties of the unfolding analysis}

The same systematic uncertainties as for the fragmentation function
fits are evaluated also for the $\mxE$ measurement, with the exception
of fit range effects, which are specific to the model-dependent fit
procedure. Binning effects are not present in RUN. The resulting
systematic uncertainties are summarised in Table~\ref{tab-unfsyst}.

An additional uncertainty is introduced by the dependence of the
$\mxE$ measurement on detector and acceptance modelling in the Monte
Carlo simulation.  As in all unfolding problems, one needs the
resolution (or spectral) function $g(x_0,x)$ where $x_0$ is the energy
of the hadrons entering the detector and $x$ their measured energy.
This function is not measured, but calculated by the OPAL detector
simulation~\cite{bib-gopal}. Since the detector simulation is
generally made in the framework of some specific Monte-Carlo program
generating hadron distributions, a small residual dependency of
$g(x_0,x)$ on the particular Monte Carlo event generator remains. To
estimate the associated systematic uncertainty, the unfolding
procedure was repeated using not only the best reweighting fit result,
but also all other models as initial estimators of the true
distribution.  This study was performed independently for all
datasets.  The unfolding was also performed with the JETSET Monte
Carlo sample replaced by HERWIG 5.9 and 6.2 samples. The check using
JETSET with the Collins-Spiller parametrisation dominates the
modelling uncertainty in the negative direction, which is taken as the
largest observed deviation from the central $\mxE$ value. The
uncertainty in the positive direction is dominated by the third best
model, which in almost all datasets is the Kartvelishvili et
al.~parametrisation.  The resulting model uncertainty is found to be
$^{+0.0024}_{-0.0016}$.

The result of the unfolding procedure with the RUN algorithm is
cross-checked with the SVD-GURU method, and the difference between the
two results is assigned as systematic uncertainty. Furthermore, a
difference of similar size is observed between the spline unfolding
result of the RUN method and a binned representation of the unfolded
distribution. This difference is also included in the unfolding method
uncertainty.

\begin{table}
\begin{center}
\begin{tabular}{|l|r|}
\hline
                 & uncertainty contribution    \\
\hline
\hline
model dependence \up{6mm}
                & $^{+0.0024}_{-0.0016}$ \\

energy resolution & $\pm0.0018$            \\
energy scale      & $\pm0.0006$            \\
$\rphi$ tracking& $\pm0.0013$            \\
$z$ tracking    & $\pm0.0002$            \\
bias modeling   & $\pm0.0011$          \\
$\qb$ baryons   & $<0.0001$ \\
$\Bs$ fraction  & $\pm0.0002$ \\
$\Bss$ fraction & $\pm0.0006$            \\
$\Bss$ Q-value  & $<0.0001$     \\
$\qb$ decay multiplicity& $\pm0.0006$ \\
$\qb$ lifetime  & $\pm0.0007$            \\
$\qc$ lifetime  & $\pm0.0001$            \\
charm fragmentation    & $\pm0.0001$              \\
$\gbb$ rate     & $\pm0.0001$            \\
$\gcc$ rate     & $\pm0.0005$ \\
$\Rb$           & $<0.0001$              \\
$\Rc$           & $<0.0001$              \\
beam energy     & $\pm0.0001$            \\
binning effects & $<0.0001$      \\
unfolding method & $\pm 0.0002$      \\
\hline
total \up{6mm}  & $^{+0.0038}_{-0.0033}$ \\
\hline
\end{tabular}
\end{center}

\smcap{\label{tab-unfsyst} Summary of all contributions to the total systematic
uncertainty of the $\mxE$ measurement from the unfolding analysis.}
\end{table}

Summing all systematic uncertainties in quadrature yields a total
systematic uncertainty on $\mxE$ of $^{+0.0038}_{-0.0033}$. As
expected, the systematic uncertainty is larger than the statistical
precision.

Table~\ref{tab-binnedresult} shows a representation of the RUN result
in 20 bins in the $\xE$ range between 0.1 and 1.0. This table, along
with the full correlation matrix of statistical
(Table~\ref{tab-errormatrix}) and systematic
(Tables~\ref{tab-systmatrixp} and \ref{tab-systmatrixn}) uncertainties
can be used to compare the OPAL results with further hadronisation
models not discussed here. It has to be pointed out again that the
$\mxE$ value obtained from the binned RUN result is smaller than the
unbinned result by $\Delta\mxE=-0.0002$. This is caused by binning
effects which are reduced by using a small bin width, but cannot be
entirely avoided.

\section{Summary and discussion}

Using an unfolding technique to reduce the dependence on $\qb$ quark
hadronisation models, the mean scaled energy of weakly decaying $\B$
hadrons in $\Z$ decays is measured to be
\[
  \mxE = 0.7193 \pm 0.0016 (stat) ^{+0.0038}_{-0.0033}(syst) \ .
\]
This is the most precise available measurement of this quantity.
Consistent results are obtained using an alternative unfolding method
and from model-dependent reweighting fits.

The result obtained here is in good agreement with a recent result
from the ALEPH Collaboration~\cite{bib-bfraleph01},
$\mxE=0.716\pm0.006(stat)\pm0.006(syst)$. ALEPH uses exclusive
semileptonic $\B$ decays, leading to a smaller candidate sample and
thus a larger statistical uncertainty. Another new measurement by
SLD~\cite{bib-bfrsld02} gives a somewhat lower value:
$\mxE=0.709\pm0.003(stat)\pm 0.003(syst)\pm0.002(model)$. The
difference between the OPAL and the SLD measurements has a statistical
significance of less than $2$ standard deviations taking only the
uncorrelated uncertainties into account.  Another $\mxE$ measurement
was recently performed using inclusive $\B\to\ell$
decays~\cite{bib-bfropal00}. Modelling of the lepton energy spectrum
introduces additional systematic errors in the lepton-based analysis.
The result of $\mxE=0.709\pm0.003(stat) \pm0.003(syst)
\pm0.013(model)$ is compatible with the analysis presented here,
especially given that the result in~\cite{bib-bfropal00} is not
model-independent, but based on a Peterson et al.~parametrisation.
The LEP average result for $\mxE$ in the framework of the Peterson et
al.~model, obtained from earlier analyses~\cite{bib-bcomb}, is
$0.702\pm0.008$, again in excellent agreement with the value of
$0.7023\pm 0.0006(stat)\pm0.0019(syst)$ found in this analysis.

The best description of the data with a fragmentation function with
one free parameter is achieved with the Kartvelishvili et al.~model.
The Peterson et al.~and Collins-Spiller models produce energy
distributions which are too broad to describe the data. Similar
features have been observed by SLD and ALEPH in their recent
publications. The Bowler and Lund parametrisations, each having two
free parameters, achieve a better $\chi^2$/d.o.f. in this analysis and
are clearly compatible with the data.  The same conclusion is reached
by SLD while ALEPH did not test these models. The HERWIG cluster model
is clearly disfavoured. The main difference of the two HERWIG versions
tested in this analysis is the amount of smearing of the $\B$ hadron
direction around the initial $\qb$ quark direction.  Significant
smearing is used in the HERWIG 5.9 sample, softening the spectrum too
much. The HERWIG 6.2 sample is used without any smearing, giving an
$\xE$ distribution which is in much better agreement with the data,
but which is still too broad. Similar results are obtained by SLD.

The fitted values of the parameters describing each hadronisation
model agree less well between the different experiments than the
measured $\mxE$ values.  The parameter values depend critically on
details of the Monte Carlo tuning, which is not identical in all
respects among the collaborations, although efforts have been made to
correct most relevant Monte Carlo parameters to a common set of
values.

A general conclusion of the analysis presented here is that the parton
shower plus string hadronisation Monte Carlo models provide a good
description of the current data. The fragmentation functions derived
from intrinsic symmetries of the string model (Bowler, Lund symmetric)
are favoured over the phenomenological approaches of Kartvelishvili et
al., Peterson et al., and Collins-Spiller.

\bigskip\bigskip
\noindent
{\large\bf Acknowledgements} \\
\par
We particularly wish to thank the SL Division for the efficient operation
of the LEP accelerator at all energies
 and for their close cooperation with
our experimental group.  In addition to the support staff at our own
institutions we are pleased to acknowledge the  \\
Department of Energy, USA, \\
National Science Foundation, USA, \\
Particle Physics and Astronomy Research Council, UK, \\
Natural Sciences and Engineering Research Council, Canada, \\
Israel Science Foundation, administered by the Israel
Academy of Science and Humanities, \\
Benoziyo Center for High Energy Physics,\\
Japanese Ministry of Education, Culture, Sports, Science and
Technology (MEXT) and a grant under the MEXT International
Science Research Program,\\
Japanese Society for the Promotion of Science (JSPS),\\
German Israeli Bi-national Science Foundation (GIF), \\
Bundesministerium f\"ur Bildung und Forschung, Germany, \\
National Research Council of Canada, \\
Hungarian Foundation for Scientific Research, OTKA T-029328, 
and T-038240,\\
Fund for Scientific Research, Flanders, F.W.O.-Vlaanderen, Belgium.\\

\clearpage

\begin{table}
\begin{center}
\begin{tabular}{|r|r|r|}
\hline
bin & \multicolumn{1}{c|}{$\xE$ range} & \multicolumn{1}{c|}{$1/\sigma\ \mathrm{d}\sigma/\mathrm{d}N$} \\
\hline
\hline
 1 & 0.100--0.145 & 0.00142 $\pm$ 0.00003 \ $^{+0.00014}_{-0.00011}$ \\
 2 & 0.145--0.190 & 0.00526 $\pm$ 0.00010 \ $^{+0.00051}_{-0.00044}$ \\
 3 & 0.190--0.235 & 0.00786 $\pm$ 0.00014 \ $^{+0.00083}_{-0.00061}$ \\
 4 & 0.235--0.280 & 0.00987 $\pm$ 0.00018 \ $^{+0.00095}_{-0.00045}$ \\
 5 & 0.280--0.325 & 0.01208 $\pm$ 0.00022 \ $^{+0.00097}_{-0.00044}$ \\
 6 & 0.325--0.370 & 0.01502 $\pm$ 0.00027 \ $^{+0.00109}_{-0.00076}$ \\
 7 & 0.370--0.415 & 0.01895 $\pm$ 0.00033 \ $^{+0.00140}_{-0.00079}$ \\
 8 & 0.415--0.460 & 0.02358 $\pm$ 0.00039 \ $^{+0.00158}_{-0.00080}$ \\
 9 & 0.460--0.505 & 0.02839 $\pm$ 0.00043 \ $^{+0.00115}_{-0.00129}$ \\
10 & 0.505--0.550 & 0.03364 $\pm$ 0.00042 \ $^{+0.00112}_{-0.00269}$ \\
11 & 0.550--0.595 & 0.04154 $\pm$ 0.00043 \ $^{+0.00222}_{-0.00380}$ \\
12 & 0.595--0.640 & 0.05469 $\pm$ 0.00047 \ $^{+0.00250}_{-0.00340}$ \\
13 & 0.640--0.685 & 0.07333 $\pm$ 0.00049 \ $^{+0.00238}_{-0.00250}$ \\
14 & 0.685--0.730 & 0.09490 $\pm$ 0.00065 \ $^{+0.00263}_{-0.00220}$ \\
15 & 0.730--0.775 & 0.11843 $\pm$ 0.00083 \ $^{+0.00211}_{-0.00212}$ \\
16 & 0.775--0.820 & 0.14007 $\pm$ 0.00069 \ $^{+0.00265}_{-0.00860}$ \\
17 & 0.820--0.865 & 0.14425 $\pm$ 0.00061 \ $^{+0.00567}_{-0.01376}$ \\
18 & 0.865--0.910 & 0.11268 $\pm$ 0.00065 \ $^{+0.00378}_{-0.00560}$ \\
19 & 0.910--0.955 & 0.05472 $\pm$ 0.00047 \ $^{+0.00391}_{-0.01055}$ \\
20 & 0.955--1.000 & 0.00933 $\pm$ 0.00012 \ $^{+0.00123}_{-0.00322}$ \\
\hline
\end{tabular}
\end{center}
\smcap{\label{tab-binnedresult} Unfolded $\xE$ distribution obtained from
the RUN program. Statistical and systematic uncertainties are given for each
 bin.
The corresponding correlation matrices are given in
Table~\ref{tab-errormatrix} (statistical uncertainties),
Table~\ref{tab-systmatrixp} (positive systematic uncertainties),
and Table~\ref{tab-systmatrixn} (negative systematic uncertainties).
A binned representation of
the RUN result will naturally lead to a slightly different $\mxE$ than
that calculated from the spline result.}
\end{table}

\begin{landscape}
\begin{table}
\small
\begin{tabular*}{23cm}{|r|*{20}{r@{\extracolsep{\fill}}}|}
\hline
bin & 1 & 2 & 3 & 4 & 5 & 6 & 7 & 8 & 9 & 10 & 11 & 12 & 13 & 14 & 15 & 16 & 17 & 18 & 19 & 20\, \\
\hline
\hline
1 & 1.00 & 1.00 & 1.00 & 1.00 & 1.00 & 1.00 & 1.00 & 0.99 & 0.97 & 0.91
  & 0.72 & 0.37 &-0.10 &-0.48 &-0.54 &-0.44 &-0.02 & 0.21 & 0.17 & 0.08\,\\
2 &      & 1.00 & 1.00 & 1.00 & 1.00 & 1.00 & 1.00 & 0.99 & 0.98 & 0.92
  & 0.71 & 0.37 &-0.10 &-0.48 &-0.54 &-0.44 &-0.02 & 0.21 & 0.16 & 0.08\,\\
3 &      &      & 1.00 & 1.00 & 1.00 & 1.00 & 1.00 & 0.99 & 0.97 & 0.91
  & 0.71 & 0.36 &-0.10 &-0.48 &-0.53 &-0.44 &-0.02 & 0.21 & 0.16 & 0.08\,\\
4 &      &      &      & 1.00 & 1.00 & 1.00 & 1.00 & 0.99 & 0.97 & 0.91
  & 0.71 & 0.37 &-0.10 &-0.48 &-0.54 &-0.44 &-0.02 & 0.21 & 0.17 & 0.08\,\\
5 &      &      &      &      & 1.00 & 1.00 & 1.00 & 0.99 & 0.98 & 0.92
  & 0.72 & 0.37 &-0.09 &-0.48 &-0.54 &-0.44 &-0.03 & 0.21 & 0.17 & 0.09\,\\
6 &      &      &      &      &      & 1.00 & 1.00 & 0.99 & 0.98 & 0.92
  & 0.72 & 0.38 &-0.09 &-0.48 &-0.54 &-0.45 &-0.03 & 0.21 & 0.17 & 0.08\,\\
7 &      &      &      &      &      &      & 1.00 & 1.00 & 0.99 & 0.93
  & 0.73 & 0.38 &-0.09 &-0.48 &-0.54 &-0.44 &-0.02 & 0.20 & 0.16 & 0.07\,\\
8 &      &      &      &      &      &      &      & 1.00 & 0.99 & 0.94
  & 0.75 & 0.40 &-0.08 &-0.47 &-0.54 &-0.45 &-0.03 & 0.20 & 0.15 & 0.07\,\\
9 &      &      &      &      &      &      &      &      & 1.00 & 0.97
  & 0.81 & 0.48 &-0.00 &-0.45 &-0.56 &-0.49 &-0.06 & 0.20 & 0.17 & 0.08\,\\
10&      &      &      &      &      &      &      &      &      & 1.00
  & 0.92 & 0.67 & 0.19 &-0.39 &-0.60 &-0.60 &-0.15 & 0.19 & 0.22 & 0.15\,\\
11& \multicolumn{10}{c}{ }
  & 1.00 & 0.90 & 0.49 &-0.20 &-0.57 &-0.69 &-0.30 & 0.13 & 0.26 & 0.25\,\\
12& \multicolumn{10}{c}{ }
  &      & 1.00 & 0.80 & 0.14 &-0.33 &-0.60 &-0.46 &-0.05 & 0.18 & 0.26\,\\
13& \multicolumn{10}{c}{ }
  &      &      & 1.00 & 0.69 & 0.26 &-0.18 &-0.57 &-0.43 &-0.13 & 0.06\,\\
14& \multicolumn{10}{c}{ }
  &      &      &      & 1.00 & 0.87 & 0.49 &-0.37 &-0.69 &-0.51 &-0.28\,\\
15& \multicolumn{10}{c}{ }
  &      &      &      &      & 1.00 & 0.83 &-0.02 &-0.58 &-0.61 &-0.46\,\\
16& \multicolumn{10}{c}{ }
  &      &      &      &      &      & 1.00 & 0.52 &-0.16 &-0.46 &-0.51\,\\
17& \multicolumn{10}{c}{ }
  &      &      &      &      &      &      & 1.00 & 0.72 & 0.26 &-0.05\,\\
18& \multicolumn{10}{c}{ }
  &      &      &      &      &      &      &      & 1.00 & 0.83 & 0.57\,\\
19& \multicolumn{10}{c}{ }
  &      &      &      &      &      &      &      &      & 1.00 & 0.93\,\\
20& \multicolumn{19}{c}{ } & 1.00\,\\
\hline
\end{tabular*}
\normalsize
\smcap{\label{tab-errormatrix} Correlation matrix of statistical uncertainties of the distribution in Table~\ref{tab-binnedresult}.}
\end{table}
\end{landscape}

\begin{landscape}
\begin{table}
\small
\begin{tabular*}{23cm}{|r|*{20}{r@{\extracolsep{\fill}}}|}
\hline
bin & 1 & 2 & 3 & 4 & 5 & 6 & 7 & 8 & 9 & 10 & 11 & 12 & 13 & 14 & 15 & 16 & 17 & 18 & 19 & 20\, \\
\hline
\hline
 1 & 1.00 & 1.00 & 1.00 & 1.00 & 0.99 & 0.99 & 0.99 & 0.98 & 0.89 &-0.28 &-0.69 &-0.67 &-0.50 &-0.51 &-0.65 & 0.17 & 0.44 & 0.18 &-0.94 &-0.74\,\\
 2 &      & 1.00 & 1.00 & 1.00 & 1.00 & 0.99 & 0.99 & 0.99 & 0.90 &-0.28 &-0.69 &-0.67 &-0.51 &-0.52 &-0.66 & 0.17 & 0.44 & 0.18 &-0.94 &-0.74\,\\
 3 &      &      & 1.00 & 1.00 & 0.99 & 0.99 & 0.99 & 0.98 & 0.88 &-0.32 &-0.72 &-0.70 &-0.54 &-0.56 &-0.67 & 0.21 & 0.48 & 0.23 &-0.95 &-0.77\,\\
 4 &      &      &      & 1.00 & 1.00 & 0.99 & 0.99 & 0.99 & 0.89 &-0.29 &-0.70 &-0.68 &-0.53 &-0.55 &-0.68 & 0.18 & 0.46 & 0.21 &-0.94 &-0.76\,\\
 5 &      &      &      &      & 1.00 & 1.00 & 1.00 & 1.00 & 0.92 &-0.22 &-0.64 &-0.63 &-0.48 &-0.52 &-0.68 & 0.11 & 0.40 & 0.15 &-0.92 &-0.71\,\\
 6 &      &      &      &      &      & 1.00 & 1.00 & 1.00 & 0.94 &-0.17 &-0.61 &-0.60 &-0.45 &-0.50 &-0.69 & 0.06 & 0.37 & 0.13 &-0.90 &-0.69\,\\
 7 &      &      &      &      &      &      & 1.00 & 1.00 & 0.93 &-0.19 &-0.63 &-0.62 &-0.48 &-0.53 &-0.71 & 0.09 & 0.40 & 0.16 &-0.91 &-0.71\,\\
 8 &      &      &      &      &      &      &      & 1.00 & 0.95 &-0.16 &-0.60 &-0.59 &-0.46 &-0.52 &-0.72 & 0.05 & 0.38 & 0.14 &-0.89 &-0.69\,\\
 9 &      &      &      &      &      &      &      &      & 1.00 & 0.17 &-0.31 &-0.31 &-0.20 &-0.34 &-0.68 &-0.27 & 0.09 &-0.11 &-0.71 &-0.44\,\\
10 &      &      &      &      &      &      &      &      &      & 1.00 & 0.88 & 0.86 & 0.78 & 0.57 & 0.13 &-0.98 &-0.88 &-0.77 & 0.54 & 0.77\,\\
11 &      &      &      &      &      &      &      &      &      &      & 1.00 & 0.99 & 0.87 & 0.74 & 0.47 &-0.82 &-0.91 &-0.72 & 0.86 & 0.96\,\\
12 &      &      &      &      &      &      &      &      &      &      &      & 1.00 & 0.94 & 0.83 & 0.57 &-0.83 &-0.95 &-0.80 & 0.82 & 0.98\,\\
13 &      &      &      &      &      &      &      &      &      &      &      &      & 1.00 & 0.95 & 0.69 &-0.80 &-0.98 &-0.93 & 0.62 & 0.92\,\\
14 &      &      &      &      &      &      &      &      &      &      &      &      &      & 1.00 & 0.86 &-0.60 &-0.89 &-0.89 & 0.56 & 0.85\,\\
15 &      &      &      &      &      &      &      &      &      &      &      &      &      &      & 1.00 &-0.13 &-0.56 &-0.55 & 0.54 & 0.66\,\\
16 &      &      &      &      &      &      &      &      &      &      &      &      &      &      &      & 1.00 & 0.89 & 0.83 &-0.44 &-0.74\,\\
17 &      &      &      &      &      &      &      &      &      &      &      &      &      &      &      &      & 1.00 & 0.94 &-0.62 &-0.92\,\\
18 &      &      &      &      &      &      &      &      &      &      &      &      &      &      &      &      &      & 1.00 &-0.34 &-0.75\,\\
19 &      &      &      &      &      &      &      &      &      &      &      &      &      &      &      &      &      &      & 1.00 & 0.87\,\\
20 &      &      &      &      &      &      &      &      &      &      &      &      &      &      &      &      &      &      &      & 1.00\,\\
\hline
\end{tabular*}
\normalsize
\smcap{\label{tab-systmatrixp} Correlation matrix of positive systematic uncertainties of the distribution in Table~\ref{tab-binnedresult}.}
\end{table}
\end{landscape}

\begin{landscape}
\begin{table}
\small
\begin{tabular*}{23cm}{|r|*{20}{r@{\extracolsep{\fill}}}|}
\hline
bin & 1 & 2 & 3 & 4 & 5 & 6 & 7 & 8 & 9 & 10 & 11 & 12 & 13 & 14 & 15 & 16 & 17 & 18 & 19 & 20\, \\
\hline
\hline
 1 & 1.00 & 0.99 & 0.99 & 0.94 &-0.00 &-0.40 &-0.28 &-0.10 &-0.44 &-0.70 &-0.75 &-0.69 &-0.45 &-0.24 & 0.58 & 0.75 & 0.73 & 0.49 &-0.90 &-0.82\,\\
 2 &      & 1.00 & 1.00 & 0.93 &-0.06 &-0.45 &-0.34 &-0.16 &-0.49 &-0.74 &-0.79 &-0.73 &-0.50 &-0.28 & 0.59 & 0.79 & 0.77 & 0.54 &-0.92 &-0.86\,\\
 3 &      &      & 1.00 & 0.93 &-0.05 &-0.45 &-0.33 &-0.15 &-0.48 &-0.74 &-0.79 &-0.73 &-0.51 &-0.30 & 0.58 & 0.78 & 0.77 & 0.54 &-0.92 &-0.86\,\\
 4 &      &      &      & 1.00 & 0.31 &-0.10 & 0.03 & 0.21 &-0.15 &-0.45 &-0.52 &-0.46 &-0.24 &-0.11 & 0.39 & 0.52 & 0.51 & 0.25 &-0.72 &-0.62\,\\
 5 &      &      &      &      & 1.00 & 0.92 & 0.96 & 0.98 & 0.88 & 0.70 & 0.64 & 0.67 & 0.70 & 0.51 &-0.45 &-0.65 &-0.65 &-0.76 & 0.43 & 0.54\,\\
 6 &      &      &      &      &      & 1.00 & 0.99 & 0.94 & 0.98 & 0.92 & 0.89 & 0.89 & 0.83 & 0.56 &-0.64 &-0.89 &-0.89 &-0.89 & 0.76 & 0.83\,\\
 7 &      &      &      &      &      &      & 1.00 & 0.98 & 0.98 & 0.87 & 0.83 & 0.84 & 0.79 & 0.53 &-0.63 &-0.83 &-0.82 &-0.85 & 0.67 & 0.75\,\\
 8 &      &      &      &      &      &      &      & 1.00 & 0.94 & 0.77 & 0.72 & 0.74 & 0.73 & 0.47 &-0.57 &-0.72 &-0.70 &-0.78 & 0.52 & 0.61\,\\
 9 &      &      &      &      &      &      &      &      & 1.00 & 0.95 & 0.92 & 0.92 & 0.84 & 0.55 &-0.70 &-0.92 &-0.90 &-0.89 & 0.78 & 0.84\,\\
10 &      &      &      &      &      &      &      &      &      & 1.00 & 1.00 & 0.99 & 0.85 & 0.55 &-0.74 &-1.00 &-0.98 &-0.90 & 0.94 & 0.97\,\\
11 &      &      &      &      &      &      &      &      &      &      & 1.00 & 0.99 & 0.85 & 0.57 &-0.72 &-1.00 &-0.99 &-0.89 & 0.96 & 0.98\,\\
12 &      &      &      &      &      &      &      &      &      &      &      & 1.00 & 0.92 & 0.66 &-0.64 &-0.99 &-0.99 &-0.94 & 0.92 & 0.97\,\\
13 &      &      &      &      &      &      &      &      &      &      &      &      & 1.00 & 0.89 &-0.30 &-0.85 &-0.90 &-0.99 & 0.71 & 0.83\,\\
14 &      &      &      &      &      &      &      &      &      &      &      &      &      & 1.00 & 0.15 &-0.57 &-0.67 &-0.86 & 0.44 & 0.59\,\\
15 &      &      &      &      &      &      &      &      &      &      &      &      &      &      & 1.00 & 0.71 & 0.61 & 0.38 &-0.72 &-0.65\,\\
16 &      &      &      &      &      &      &      &      &      &      &      &      &      &      &      & 1.00 & 0.99 & 0.90 &-0.96 &-0.99\,\\
17 &      &      &      &      &      &      &      &      &      &      &      &      &      &      &      &      & 1.00 & 0.94 &-0.94 &-0.99\,\\
18 &      &      &      &      &      &      &      &      &      &      &      &      &      &      &      &      &      & 1.00 &-0.77 &-0.88\,\\
19 &      &      &      &      &      &      &      &      &      &      &      &      &      &      &      &      &      &      & 1.00 & 0.98\,\\
20 &      &      &      &      &      &      &      &      &      &      &      &      &      &      &      &      &      &      &      & 1.00\,\\
\hline
\end{tabular*}
\normalsize
\smcap{\label{tab-systmatrixn} Correlation matrix of negative systematic uncertainties of the distribution in Table~\ref{tab-binnedresult}.}
\end{table}
\end{landscape}

\end{document}